\documentclass[useAMS,usenatbib]{mn2e}

\usepackage{graphicx}
\usepackage[T1]{fontenc}
\usepackage{hyperref}
\usepackage{ifthen}

\def\glc{{\sc Galacticus}}

\newcounter{CDMDone}
\def\CDM{\ifthenelse{\equal{\arabic{CDMDone}}{0}}{cold dark matter (CDM) \setcounter{CDMDone}{1}}{CDM}}

\title{Convergence of Galaxy Properties with Merger Tree Temporal Resolution}
\author[Benson et al.]{Andrew J. Benson$^1$\thanks{E-mail: {\tt abenson@caltech.edu}}, Stefano Borgani$^{2,3,4}$, Gabriella De Lucia$^3$, Michael Boylan-Kolchin$^5$ \newauthor \& Pierluigi Monaco$^{2,3}$ \\
$^1$ California Institute of Technology, MC350-17, 1200 E. California Blvd., Pasadena, CA 91125, U.S.A. \\
$^2$ Dipartimento di Fisica dell'Universit\`a, Sezione de Astronomia, via Tiepolo 11, I-34131 Trieste, Italy \\
$^3$ INAF --- Isservatorio Astronomico di Trieste, via Tiepolo 11, I-34131, Trieste, Italy \\
$^4$ INFN --- Istituto Nazionale di Fisica Nucleare, Trieste, Italy \\
$^5$ Department of Physics and Astronomy, Center for Cosmology, University of California, 4129 Reines Hall, Irvine, CA 92697, USA}

\begin{document}

\maketitle

\begin{abstract}
Dark matter halo merger trees are now routinely extracted from cosmological simulations of structure formation. These trees are frequently used as inputs to semi-analytic models of galaxy formation to provide the backbone within which galaxy formation takes place. By necessity, these merger trees are constructed from a finite set of discrete ``snapshots'' of the N-body simulation and so have a limited temporal resolution. To date, there has been little consideration of how this temporal resolution affects the properties of galaxies formed within these trees. In particular, the question of how many snapshots are needed to achieve convergence in galaxy properties has not be answered. Therefore, we study the convergence in the stellar and total baryonic masses of galaxies, distribution of merger times, stellar mass functions and star formation rates in the \glc\ model of galaxy formation as a function of the number of ``snapshot'' times used to represent dark matter halo merger trees. When utilizing snapshots between $z=20$ and $z=0$, we find that at least 128 snapshots are required to achieve convergence to within 5\% for galaxy masses. This convergence is obtained for mean quantities averaged over large samples of galaxies---significant variance for individual galaxies remains even when using very large numbers of snapshots. We find only weak dependence of the rate of convergence on the distribution of snapshots in time---snapshots spaced uniformly in the expansion factor, uniformly in the logarithm of expansion factor or uniformly in the logarithm of critical overdensity for collapse work equally well in almost all cases. We provide input parameters to \glc\ which allow this type of convergence study to be tuned to other simulations and to be carried out for other galaxy properties.
\end{abstract}


\section{Introduction}

Simulations of the cosmological evolution of large scale structure and individual galaxies in cold dark matter cosmogonies \citep{springel_simulations_2005,kuhlen_via_2008,springel_aquarius_2008,boylan-kolchin_resolving_2009,klypin_halos_2010,prada_halo_2011} are an invaluable tool in studying the formation and evolution of galaxies. With the high resolutions and large volumes attained by modern simulations, a particularly useful approach is to extract information on individual dark matter halos and how those halos merge during the process of hierarchical structure formation. It is within these merging hierarchies of dark matter halos that galaxies form. Typically, this information on hierarchical growth is encoded as a ``merger tree'' represented by a set of ``nodes'' (each corresponding to a dark matter halo at one point in its evolution) and ``branches'' connecting those nodes which link descendant halos to their progenitors (i.e. halos which will merge together to form the descendant). While these merger trees have many applications, a very common use is an input to models which aim to solve the physics of galaxy formation within the merging hierarchy of dark matter---in particular the class known as ``semi-analytic models'' \citep{cole_hierarchical_2000,hatton_galics-_2003,croton_many_2006,de_lucia_hierarchical_2007,monaco_morgana_2007,somerville_semi-analytic_2008}.

Due to the way N-body simulations are analyzed and their data stored, merger trees are usually constructed from a set of ``snapshots''---outputs of all of the particle data at a set of $N$ times, $t_i$ where $i=1$ to $N$. Since $N$ is finite (and often limited by available storage and processing power) this means that each merger tree is a temporally sparse representation of some underlying true merger tree which may have structure on shorter timescales. Such structure is lost in the sparse representation. The question of how this loss of information affects the properties of galaxies produced by models utilizing sparse merger trees has been touched upon before, but never carefully studied. For example, \cite{helly_galaxy_2003}, utilizing merger trees with 44 snapshots (spaced uniformly in the logarithm of expansion factor) from $z=0$ to $z=20$, compared their results to those obtained from their standard {\sc Galform} model based on merger trees constructed from Press-Schechter techniques using a significantly larger number of snapshots and noted that ``\ldots we find that if we degrade the time resolution of the standard {\sc Galform} model to match that of our N-body model the properties of the galaxy populations predicted change very little.'' However, this statement refers only to statistical properties of the entire galaxy population rather than to individual galaxies and, in any case, no details were given. \cite{hatton_galics-_2003} utilized merger trees with around 70 snapshots between $z=0$ and $z=10$. They repeated their calculations using only every second snapshot and found that this leads to a 20\% scatter in cold gas content in galaxies, and a 40\% scatter in the mass of hot gas associated with each galaxy. They concluded that this was not a significant problem, but this is arguably a significant discrepancy given the accuracy of modern observational datasets. \cite{croton_many_2006} describe a galaxy formation model utilizing merger trees extracted from the Millennium Simulation with 60 snapshots between $z=0$ and $z=20$ but do not discuss if this number is sufficient to achieve converged results. \cite{somerville_semi-analytic_2008} present a model based on merger trees constructed using an extended Press-Schechter algorithm and note that they find no significant changes if they instead implement their model in merger trees extracted from N-body simulations, but do not provide further details.

Importantly, none of these studies were able to address the issue of how fast galaxy properties converge as the number of snapshots is increased, or what is the optimal distribution of snapshots in time. Additionally, in all of the galaxy formation models employed at least some of the baryonic physics (e.g. the rate of gas cooling) is solved using timesteps tied to the snapshot spacing---this is non-optimal as it is in principle possible to obtain better convergence by solving the baryonic physics on a much shorter timescale (and interpolating the dark matter halo properties as necessary).

In this work, we address these issues by utilizing the \glc\ semi-analytic galaxy formation code \citep{benson_galacticus:_2010}. This code has two key features which make it ideally suited to this problem. The first is its ability to construct merger trees using a modified extended Press-Schechter formalism \citep{parkinson_generating_2008} which both agree with the statistics of merger trees extracted from N-body simulations and which can be built without reference to any fixed grid of snapshot times, thereby achieving arbitrarily high temporal resolution. As we will describe below, \glc\ is able to post-process these trees to construct sparse representations in the same manner as trees extracted from an N-body simulation. The second key feature of \glc\ is that all baryonic physics is solved using adaptive timesteps which are adjusted to keep a specified tolerance in the quantities being computed. As such, the number and distribution of snapshots in the merger trees affects the results from \glc\ only due to the inevitable information loss resulting from the sparse representation. We exploit these features to explore convergence of basic galaxy properties as a function of the number of snapshots available and the distribution of those snapshots in time. 

The remainder of this paper is arranged as follows. In \S\ref{sec:Method} we describe the process by which we construct merger trees and alter their temporal resolution. In \S\ref{sec:Results} we present results from our convergence study and in \S\ref{sec:Conclusions} we give our conclusions.

\section{Method}\label{sec:Method}

We build merger trees using \glc's modified extended Press-Schechter algorithm\footnote{Full details can be found in the \protect\glc\ manual available online at \href{http://www.ctcp.caltech.edu/galacticus/Galacticus_v0.9.0.pdf}{\tt http://www.ctcp.caltech.edu/galacticus/Galacticus\_v0.9.0.pdf}.} \citep{parkinson_generating_2008,benson_galacticus:_2010}. This algorithm makes no reference to any fixed timesteps, but instead steps along each branch of the tree using a step that is controlled by three parameters. The first parameter, called $\epsilon_1$ by \cite{parkinson_generating_2008} and corresponding to the input parameter {\tt [modifiedPressSchechterFirstOrderAccuracy]} in \glc, ensures that the merger rate equation derived by \cite{parkinson_generating_2008} is first order accurate. The second parameter, called $\epsilon_2$ by \cite{parkinson_generating_2008} and corresponding to the input parameter {\tt [mergerTreeBuildCole2000MergeProbability]} in \glc, limits the probability that an above-resolution binary split occurs during the step. The final parameter, which we will label $\epsilon_3$ and which corresponds to the input parameter {\tt [mergerTreeBuildCole2000AccretionLimit]} in \glc, limits the maximum fractional amount of smooth accretion (i.e. sub-resolution merging) over the step\footnote{\protect\cite{parkinson_generating_2008} do not include this constraint when setting timesteps during tree building. We include it to ensure that the mass vs. time relation is sufficiently well resolved along each branch in the regime of smooth accretion.}. We typically set the values of these parameter to $0.1$ (as do \cite{parkinson_generating_2008} for $\epsilon_1$ and $\epsilon_2$) but have checked that reducing them by an order of magnitude does not affect the galaxy properties that result when using these trees (see \S\ref{app:converge}). 

\begin{figure*}
 \begin{center}
  \begin{tabular}{cc}
   \includegraphics[height=180mm]{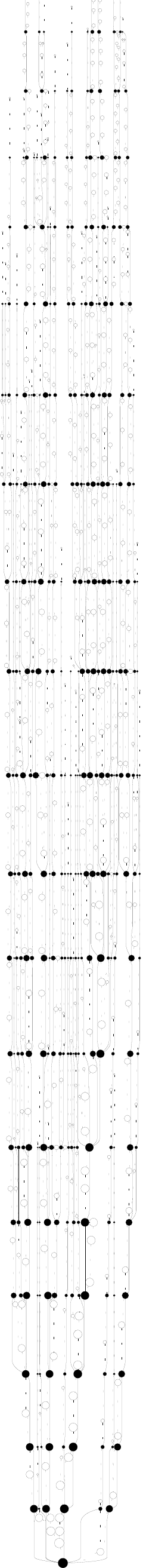}\hspace{40mm} &
   \includegraphics[trim=2mm 0mm 2mm 218mm, clip=true, width=80mm]{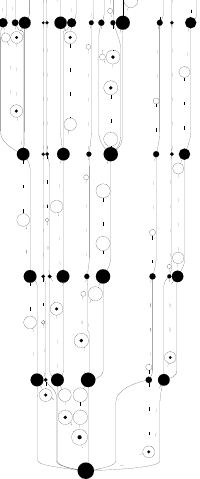}
  \end{tabular}
 \end{center}
 \caption{An example of a merger tree interpolated onto a fixed grid of timesteps. Circles represent nodes (halos) in the merger tree, with radius proportional to the logarithm of halo mass. Logarithmic time runs down the page with $z=0$ at the bottom of the figure. Open symbols and dotted lines represent nodes in the original tree and their connecting branches, while filled symbols and solid lines show nodes in the re-gridded tree and their connecting branches. The left view shows the full tree, while the right view shows a zoom in to the final five snapshot times in the same tree. In the zoomed view, nodes in the original tree which have two progenitors (i.e. are the result of a merger) are indicated by a filled dot in the center of the corresponding open circle.}
 \label{fig:tree}
\end{figure*}

Having built trees, we use the \glc\ task {\tt mergerTreeRegridTimes} to force the trees onto a pre-specified time grid---a process that we will refer to as ``re-gridding''. This process works as follows. We begin by defining a set of times, $t_i$ with $i=1$ to $N$, which would correspond to the snapshots of an N-body simulation. Given a merger tree, we walk along each branch and look for branches (i.e. connections between nodes) which span one or more of these times. When such a branch is found, we interpolate the mass of the halo along the branch if this is a primary progenitor (otherwise we keep the mass of the halo fixed along the branch) to each intersected time and create a new node at that time with the interpolated mass\footnote{When constructing trees using a modified extended Press-Schechter algorithm the primary progenitor is identified as the most massive progenitor at each bifurcation of the merger tree. When solving galaxy formation physics, \protect\glc\ performs the same interpolation of halo mass along each branch to provide a smooth evolution. Thus, this interpolation of the merger trees is consistent with the way in which galaxies and halos are evolved in \protect\glc.}. Having done this for every branch of the tree we then walk the tree again removing all nodes which do not fall on one of the snapshot times. When removing a node, we connect any children to the node's parent\footnote{By ``children'' we mean direct progenitors of a halo existing at an earlier time and by ``parent'' we mean the halo into which the halo in question will merge in the future.}. In this way we preserve the merging structure of the tree. Figure~\ref{fig:tree} shows a representation of a typical merger tree re-gridded in this way. Circles represent nodes in the tree with lines showing the branches. Circle size is proportional to the logarithm of halo mass and time runs down the page on a logarithmic axis (such that the present day is at the bottom of the figure). Open symbols show nodes in the original merger tree, built without reference to any set of snapshot times. Dotted lines show the branches in this tree. Filled circles show interpolated nodes added to the tree at each snapshot time, with solid lines indicating the branches in the resulting sparse tree. Note that in the original tree only binary mergers occur by construction (indicated by filled dots in the centres of open circles in the right-hand panel), while in the sparse tree non-binary mergers can occur.

\begin{figure}
 \begin{center}
 \includegraphics[width=80mm]{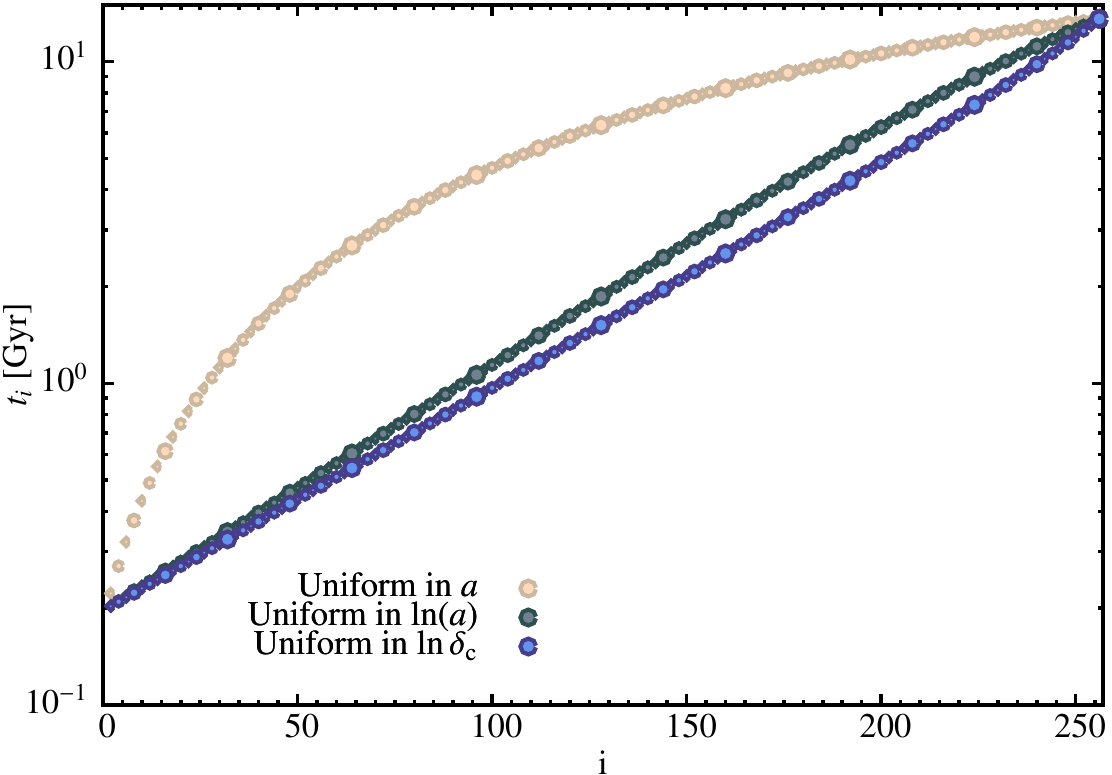}
 \end{center}
 \caption{The snapshot times used in this work when beginning snapshots at $z=20$. Colours correspond to different snapshot distribution choices as indicated in the figure. Larger points correspond to the snapshots used in coarser re-griddings.}
 \label{fig:snapshots}
\end{figure}

We consider three different options for setting the snapshot times. In each case, $t_1$ corresponds to $1+z=20$ while $t_N$ corresponds to $1+z=1$. We then consider cases where snapshots are uniformly spaced in expansion factor, $a$, are uniformly spaced in the logarithm of expansion factor, $\ln(a)$, and are uniformly spaced in the logarithm of the critical linear theory overdensity for collapse, $\ln\delta_{\rm c}$, as this is a natural timescale for structure formation (see \citealt{benson_self-consistent_2005}). Examples of the resulting distribution of snapshot times are shown in Fig.~\ref{fig:snapshots}. Note that the ``uniform in $\ln a$'' and ``uniform in $\ln\delta_{\rm c}$'' distributions are quite similar due to the fact that $\delta_{\rm c} \propto 1/a$ at high redshifts. In each case we consider $N=16$, 32, 64, 128 and 256 to explore increasingly well-resolved trees.

For each set of snapshot times we use \glc\ to compute the properties of galaxies forming in a large number of realizations of merger trees spanning the mass range $10^{11}$ to $10^{14}M_\odot$. In each case we also run a model in which no re-gridding of the merger trees occurs---we will use this as our reference point corresponding to $N \rightarrow \infty$. For specificity we adopt cosmological parameters corresponding to the WMAP-7 dataset 
($\Omega_0 = 0.2725$, $\Omega_\Lambda = 0.7275$, $\Omega_{\rm b} = 0.0455$, $\sigma_8 = 0.807$, $H_0 = 70.2$, $n_{\rm s} = 0.961$, $N_{\rm eff} = 4.34$; \citealt{komatsu_seven-year_2010}), use the \cite{eisenstein_power_1999} fitting formula for the cold dark matter transfer function and a merger tree mass resolution (i.e. the lowest mass halo which is tracked in the merger tree) of $1.2\times 10^9M_\odot$. This specific mass resolution was chosen to match that used in work currently in preparation, but corresponds to a moderately well resolved halo (i.e. a little over 100 particles) in the recent Millennium-II simulation \citep{boylan-kolchin_resolving_2009}. We consider two sets of input parameters to the \glc\ model. In the first, which we will refer to as the ``full'' model, we employ the full set of galaxy formation physics (see \citealt{benson_galacticus:_2010}) with parameters chosen to provide good fits (Benson 2011, in preparation) to the local galaxy stellar mass function \citep{li_distribution_2009}, the local galaxy HI mass function \citep{zwaan_hipass_2005} and the star formation history of the Universe \citep{hopkins_evolution_2004}. For the second set of input parameters we consider a simplified model, which we refer to as the ``simple'' model, in which we switch off all feedback effects due to supernovae and AGN (and thereby prevent any outflows of mass from galaxies) but leave all other parameters unchanged.

\subsection{Convergence in Tree Building and Baryonic Solver}\label{app:converge}

To reliably test the convergence of galaxy properties with respect to the temporal resolution of merger trees we must first confirm that the galaxy properties are converged with respect to other numerical parameters in \glc. In particular, there are two separate issues which must be considered. The first issue is the numerical accuracy with which the $N_{\rm step}\rightarrow\infty$ trees are built. The second issue is the accuracy with which the ordinary differential equations (ODEs) describing baryonic physics are solved.

\begin{figure*}
 \begin{center}
  \begin{tabular}{cc}
   \includegraphics[height=60mm]{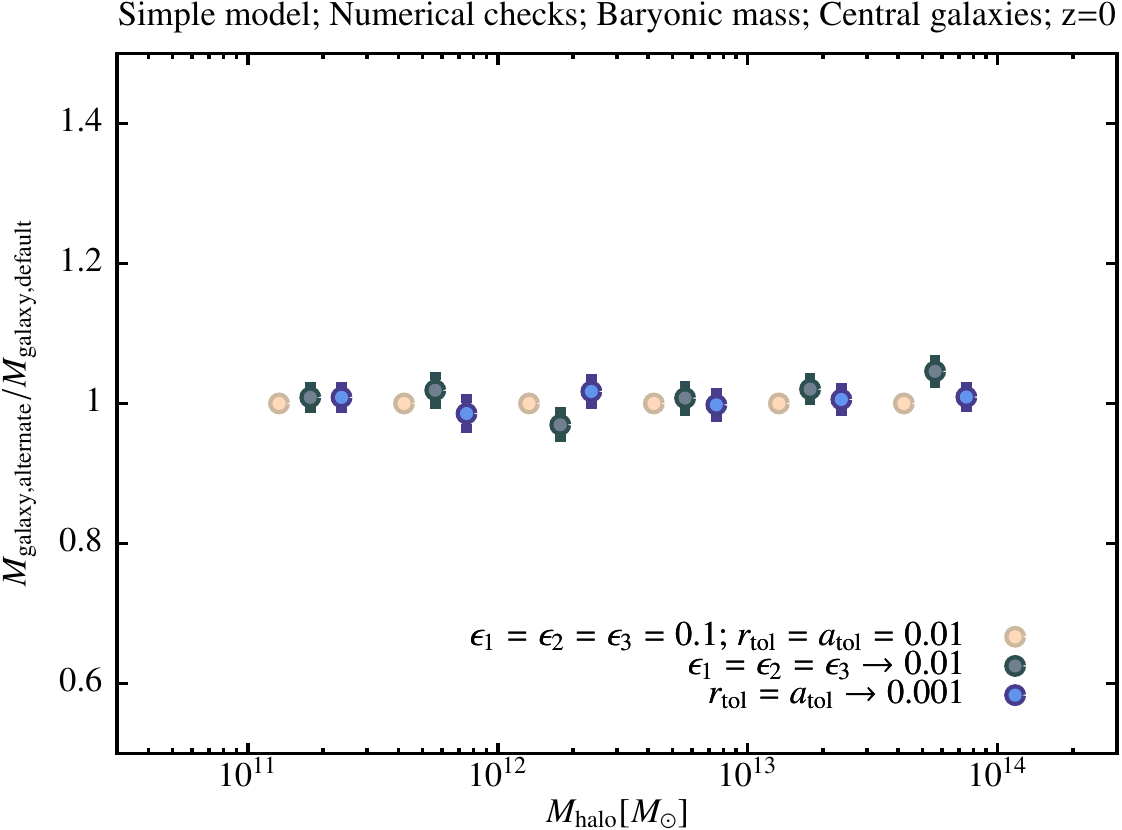} &  \includegraphics[height=60mm]{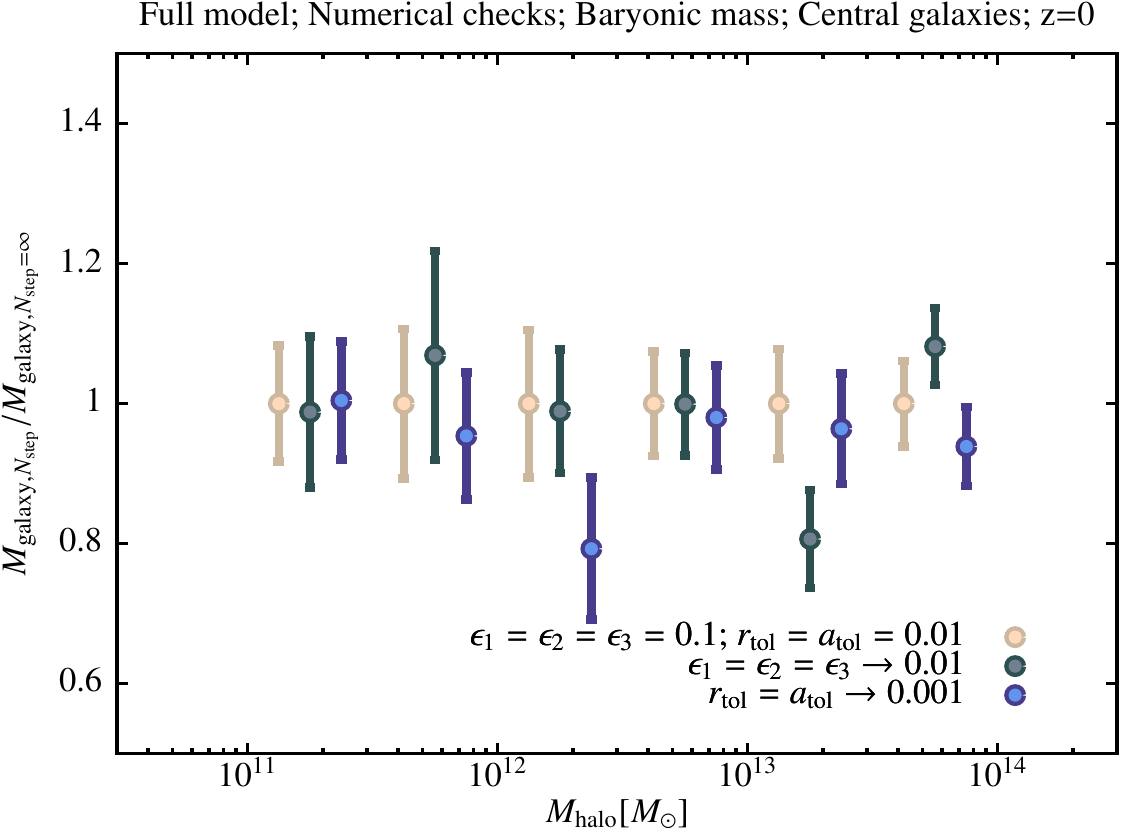} \\
   \includegraphics[height=60mm]{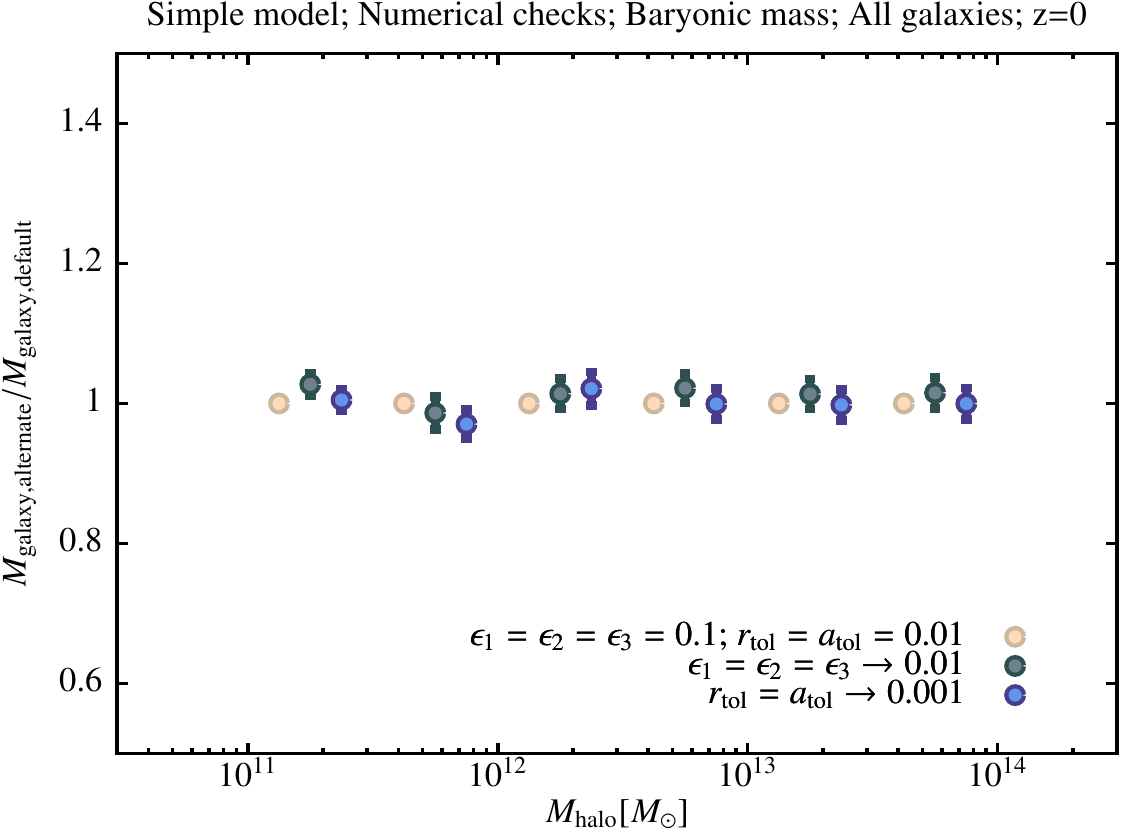} &  \includegraphics[height=60mm]{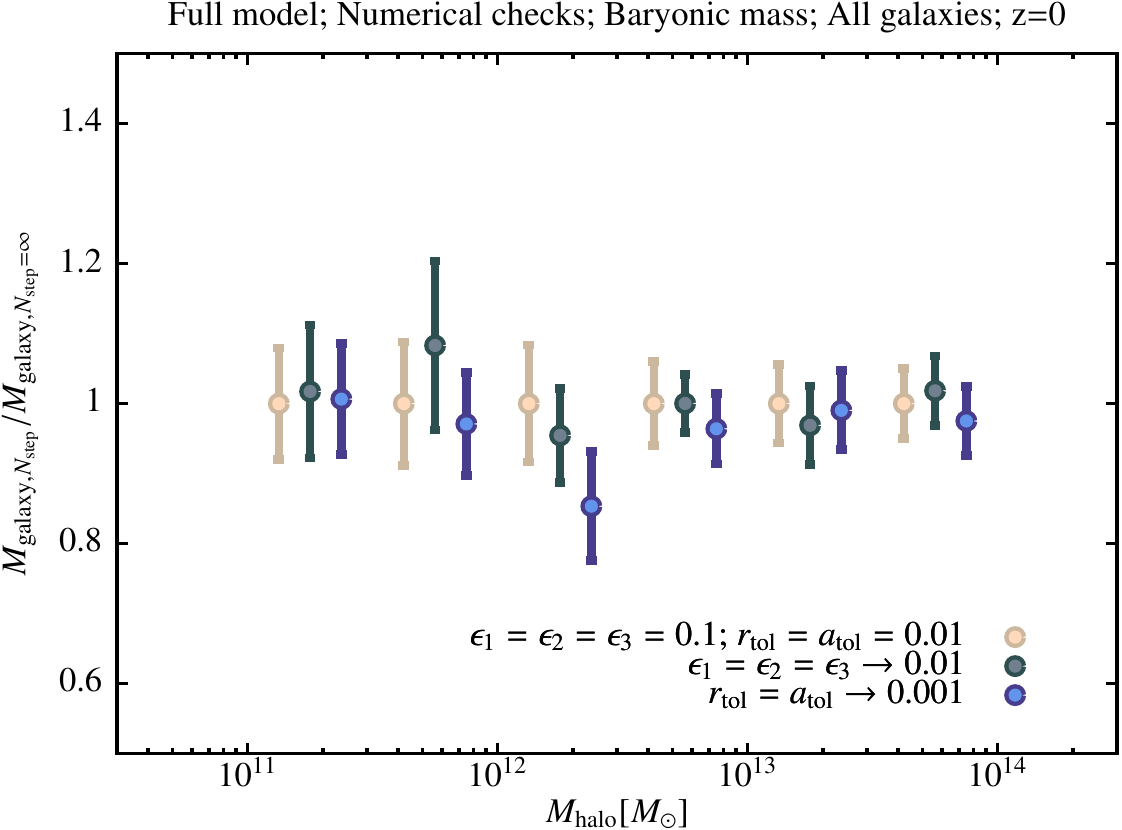} \\
   \includegraphics[height=60mm]{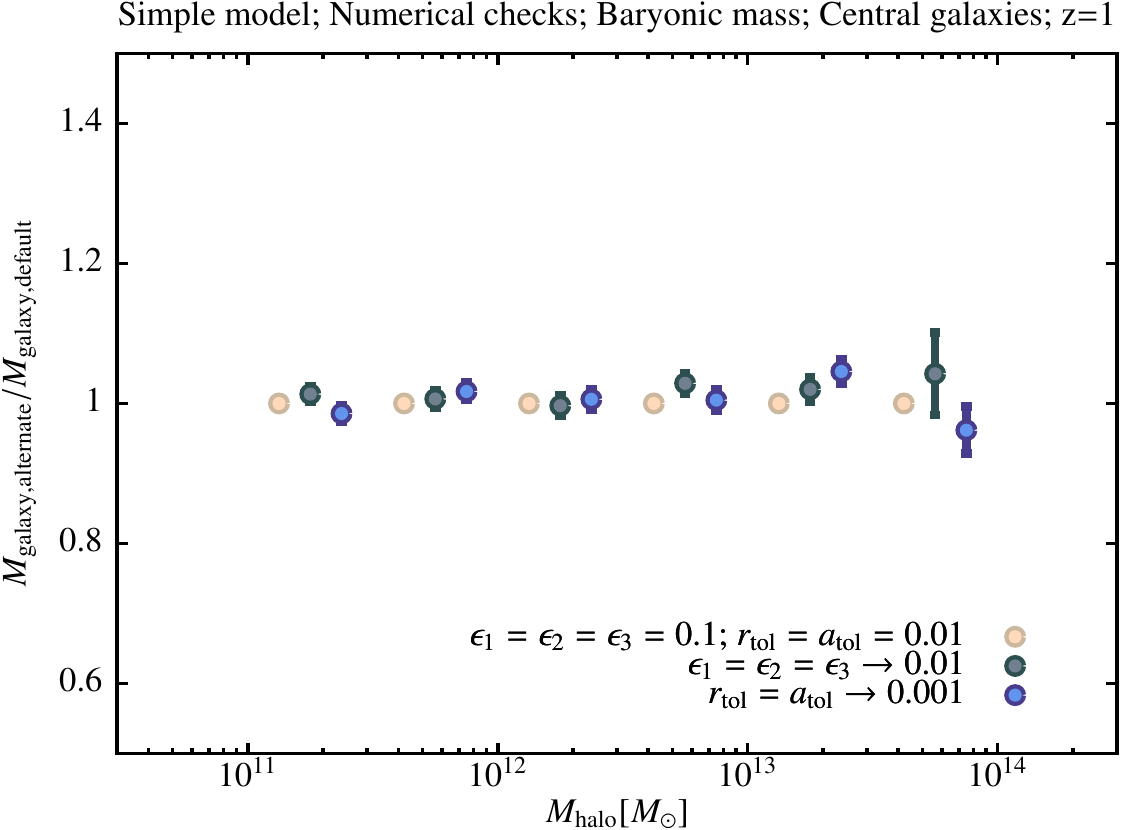} &  \includegraphics[height=60mm]{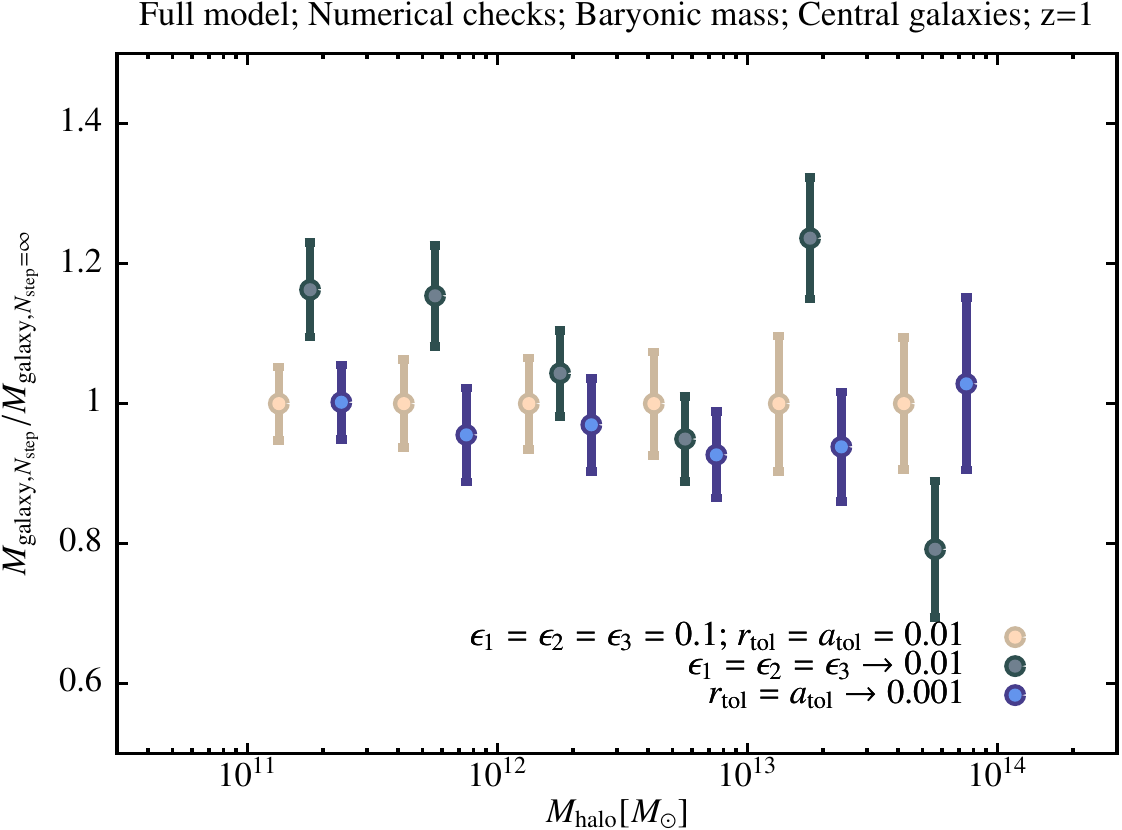} \\
  \end{tabular}
 \end{center}
 \caption{Convergence of the mean total baryonic mass of galaxies as a function of halo mass with respect to other numerical parameters in \protect\glc. Panels in the left column correspond to the simple model, while those in the right column correspond to the full model. Rows correspond to central galaxies at $z=0$, all galaxies at $z=0$ and central galaxies at $z=1$ from top to bottom. In all cases, the earliest snapshot is at $1+z=20$ and the final snapshot at $1+z=1$. Symbol colour corresponds to the parameter values used, while error bars indicate the error on the mean mass due to the finite number of merger trees realized in each bin. Points in each mass bin are given small horizontal offsets for clarity. The parameters $\epsilon_1$, $\epsilon_2$ and $\epsilon_3$ control the accuracy with which the $N_{\rm step}=\infty$ trees are built. Reducing their values results in more accurate tree construction and clearly shows that the $N_{\rm step}=\infty$ trees are well converged with respect to these parameters. The parameters $r_{\rm rol}$ and $a_{\rm tol}$ (corresponding to input parameters {\tt [odeToleranceRelative]} and {\tt [odeToleranceAbsolute]} in \glc) control the accuracy with which baryonic physics is solved in \protect\glc. Again, reducing their values results in more accurate solutions and shows that the results are sufficiently converged for this study.}
 \label{fig:convergenceOther}
\end{figure*}

Figure~\ref{fig:convergenceOther} demonstrates the convergence of galaxy properties with parameters controlling these aspects of \glc\ in $N_{\rm step}=\infty$ merger trees. We compare our standard calculation, computed with $\epsilon_1=\epsilon_2=\epsilon_3=0.1$ (which control the accuracy of merger tree construction) and $r_{\rm tol}=a_{\rm tol}=0.01$ (which control the tolerance in the ODE solver used for evolving baryonic physics and which correspond to input parameters {\tt [odeToleranceRelative]} and {\tt [odeToleranceAbsolute]} in \glc) with calculations in which these two sets of parameters are decreased by a factor of 10. For the simple model, the results are clearly extremely well converged with respect to these parameters. For the full model the scatter on the mean values are larger than in the simple model. This is not surprising as the additional physics of outflows included in the full model leads to a greater dependence of baryonic properties on details of the tree formation history. Nevertheless, given the size of the error bars the results are consistent with the trees and ODE solver being converged with respect to these numerical parameters. One possible marginal exception is the case of the baryonic mass\footnote{Following the usual logic of semi-analytic models, the baryonic mass of each galaxy is taken to be the mass of stars plus any mass in cold interstellar medium gas. Hot gas (with a temperature approximately equal to the virial temperature of the halo) and gas in cold flows is instead associated with the halo and is not included in the mass of the galaxy as discussed here.} of central galaxies at $z=1$ in the full model---there are a few points which deviate significantly, which could suggest that the trees are not sufficiently accurate for this model at high-$z$. The effect is marginal, but should be kept in mind when considering convergence in properties of high-$z$ galaxies in the full model.

\section{Results}\label{sec:Results}

We now compare results obtained from $N_{\rm step}=\infty$ trees with those from trees with finite $N_{\rm step}$ and assess the degree of convergence as a function of $N_{\rm step}$.

\subsection{Individual Galaxies}

\begin{figure*}
 \begin{center}
  \begin{tabular}{cc}
   \includegraphics[height=60mm]{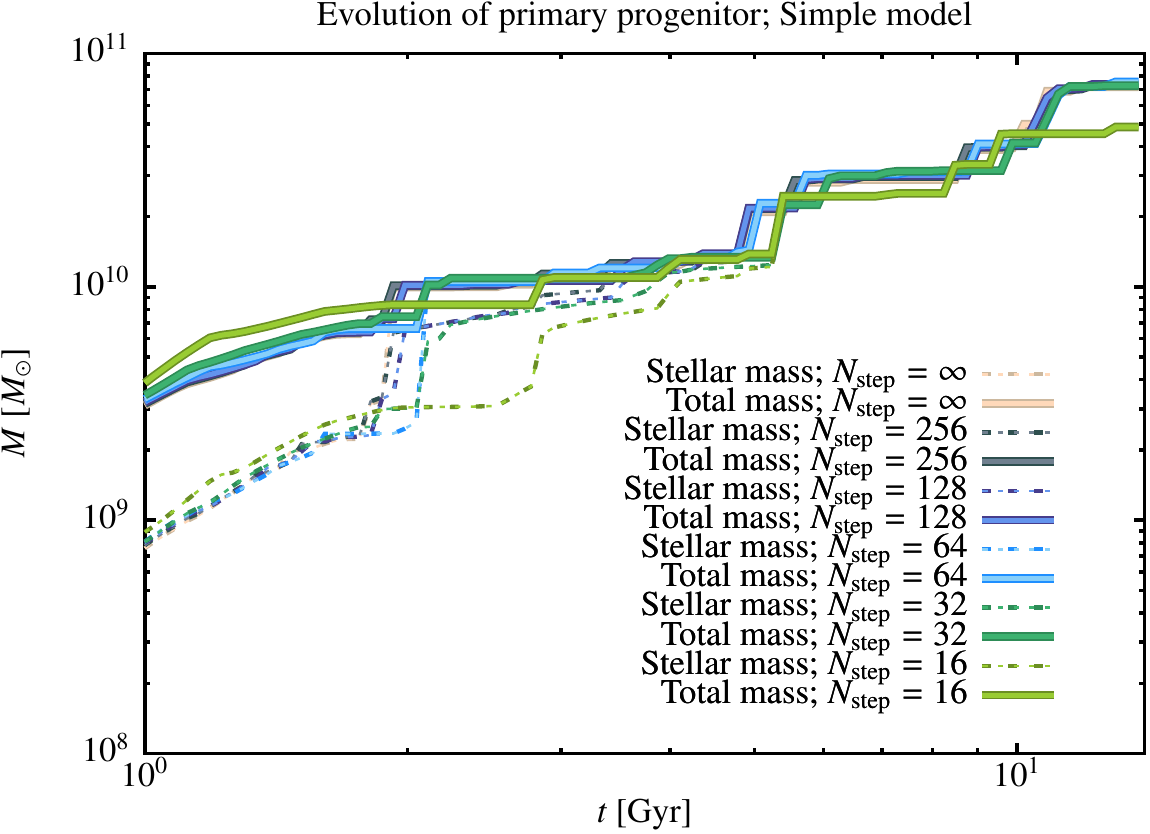} &  \includegraphics[height=60mm]{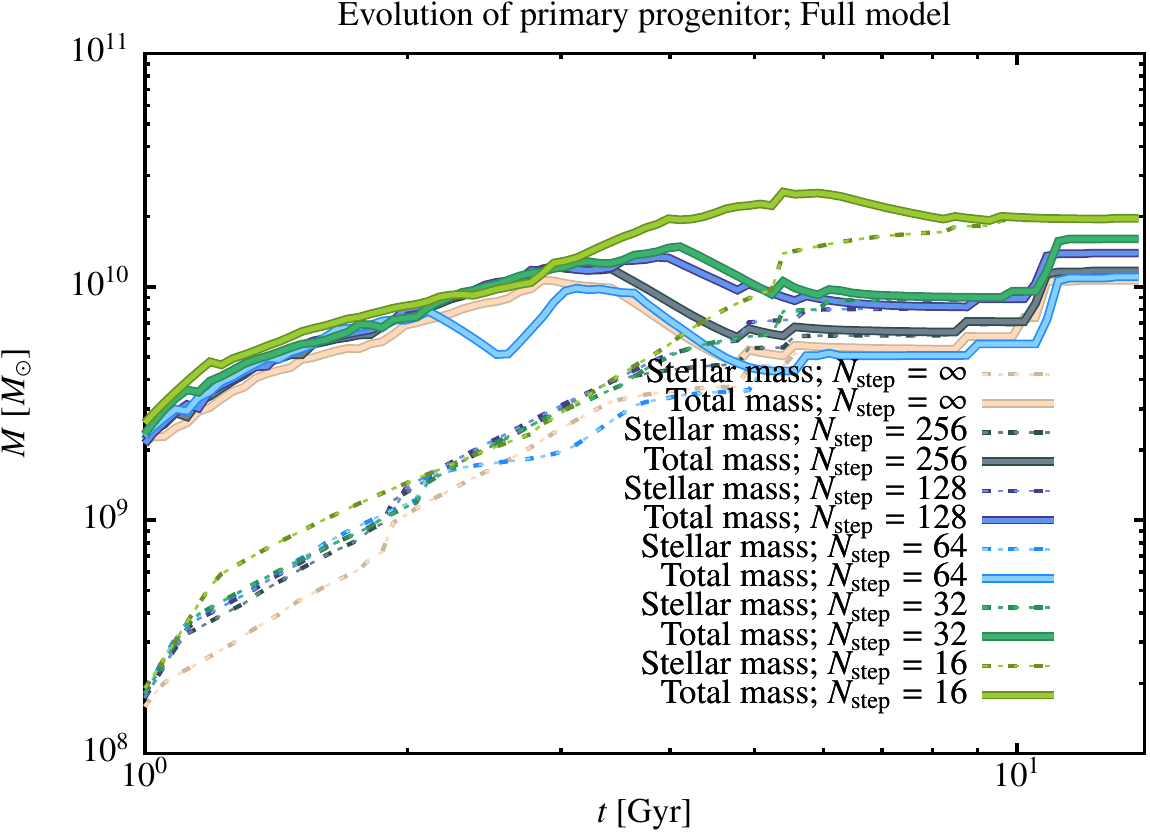} \\
   \includegraphics[height=60mm]{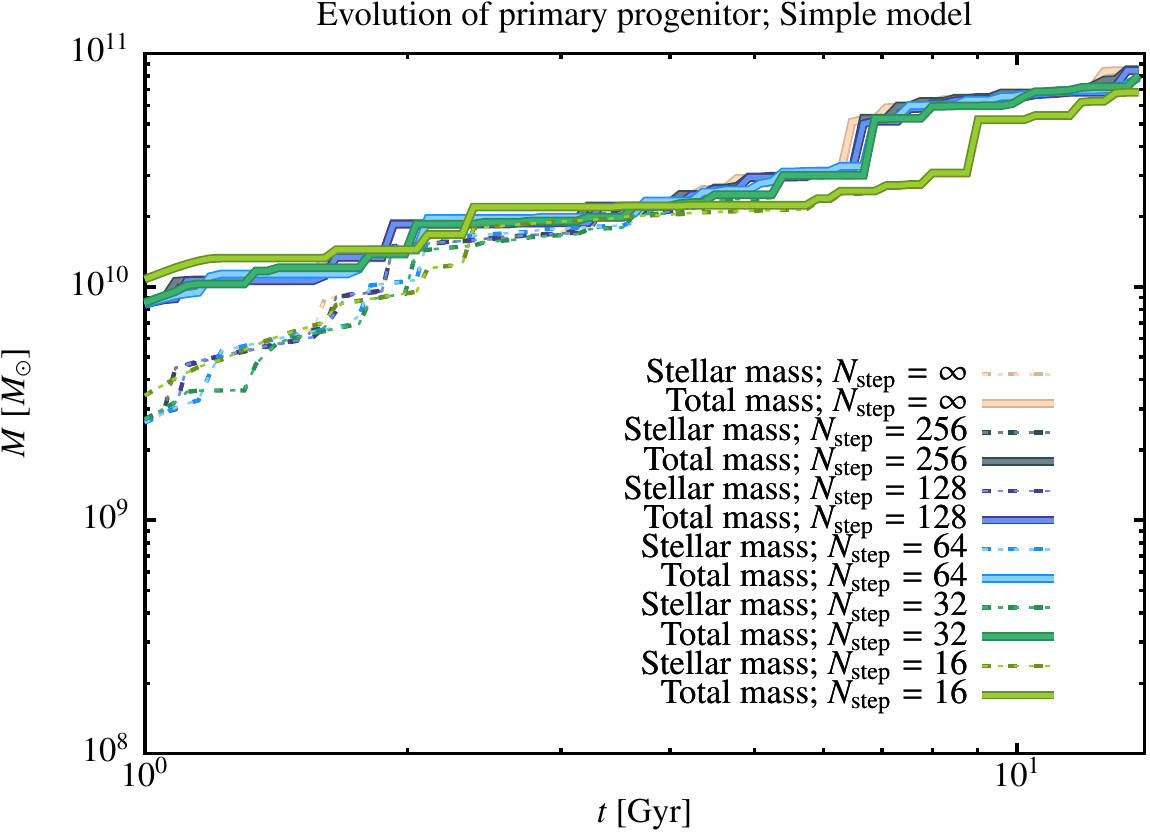} &  \includegraphics[height=60mm]{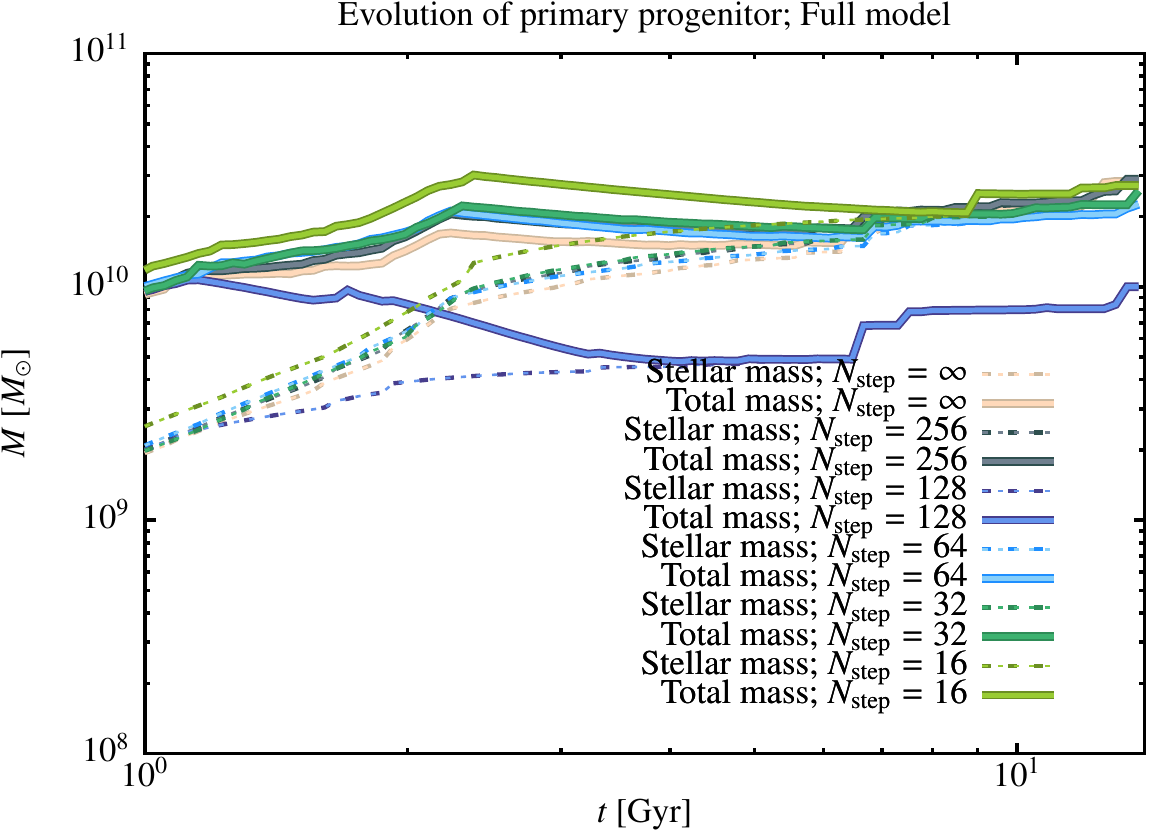}
  \end{tabular}
 \end{center}
 \caption{The evolution of the stellar (dashed lines) and total baryonic (solid lines) mass in the main progenitor galaxy in representative merger trees is shown from $z\approx 6$ to $z=0$. Colours indicate the number of snapshot times used in each calculation. The left column shows results for the simple model, while the right panel shows results for the full model.}
 \label{fig:convergenceIndividuals}
\end{figure*}

We begin by considering the evolution of individual galaxies, chosen to be the main progenitors of the central galaxy of each $z=0$ halo. Figure~\ref{fig:convergenceIndividuals} shows the stellar and total baryonic masses as a function of time for two representative galaxies (corresponding to $z=0$ halos with masses in the range $10^{12}$--$10^{13}M_\odot$) from the simple and full models. Line colours indicate the different values of $N_{\rm step}$ used. Considering the simple model first, it is clear that convergence is reached quite rapidly---while the evolution for $N_{\rm step}=16$ is significantly different from the $N_{\rm step}=\infty$ case, $N_{\rm step}=32$ is sufficient to reproduce the evolution with reasonable accuracy and $N_{\rm step} \ge 128$ results in a faithful recreation of the $N_{\rm step}=\infty$ result.

The full model shows a rather different behaviour. Convergence is generally slower and, more importantly, there are cases where even for $N_{\rm step}=256$ the evolution can diverge substantially from the $N_{\rm step}=\infty$ case. The reason for this is that the evolution of a given galaxy can depend crucially on individual events. For example, a change in the structure of the merger tree due to regridding could cause a galaxy merger event to change from being considered ``minor'' to being ``major''. Given the rules adopted by semi-analytic models this can result in very different properties for the galaxy emerging from the merger event and, therefore, to dramatically different evolution at later times. This can be seen clearly in the lower right panel of Fig.~\ref{fig:convergenceIndividuals} in which the $N_{\rm step}=256$ case diverges from the other models very early and consequently differs by a factor of $3$ in mass compared to the $N_{\rm step}=\infty$ case at $z=0$, while even the $N_{\rm step}=16$ model gets very close to the evolution and final mass of the $N_{\rm step}=\infty$ case.

Fortunately, semi-analytic models are not usually in the business of predicting the properties of individual galaxies\footnote{One exception to this is the study of \protect\cite{stringer_analytic_2010} who follow the formation of a galaxy in a merger tree extracted from an N-body+hydrodynamical simulation and compare its properties and evolution to the \emph{same} galaxy found in the \emph{same individual merger tree} in the simulation. That study used 49 timesteps and found very good agreement between the semi-analytic and simulated galaxy properties. Based on the results obtained here, we would predict that for a significant fraction of trees similar studies would find quite divergent results between the two techniques. Another class of exception is in applying semi-analytic techniques to high resolution simulations of individual halos, such as the Aquarius \protect\citep{springel_aquarius_2008} and Via Lactea \protect\citep{kuhlen_via_2008} simulations.}, but rather in making statistical predictions for populations of galaxies. Divergences due to temporal regridding such as are seen in Fig.~\ref{fig:convergenceIndividuals} should cancel to some extent when averaged over many galaxies. In the following sections we will assess how effective this averaging is by considering the convergence of average properties of galaxies.

\subsection{Statistical Comparison}

\subsubsection{Stellar and Baryonic Masses}

\begin{figure*}
 \begin{center}
  \begin{tabular}{cc}
   \includegraphics[height=60mm]{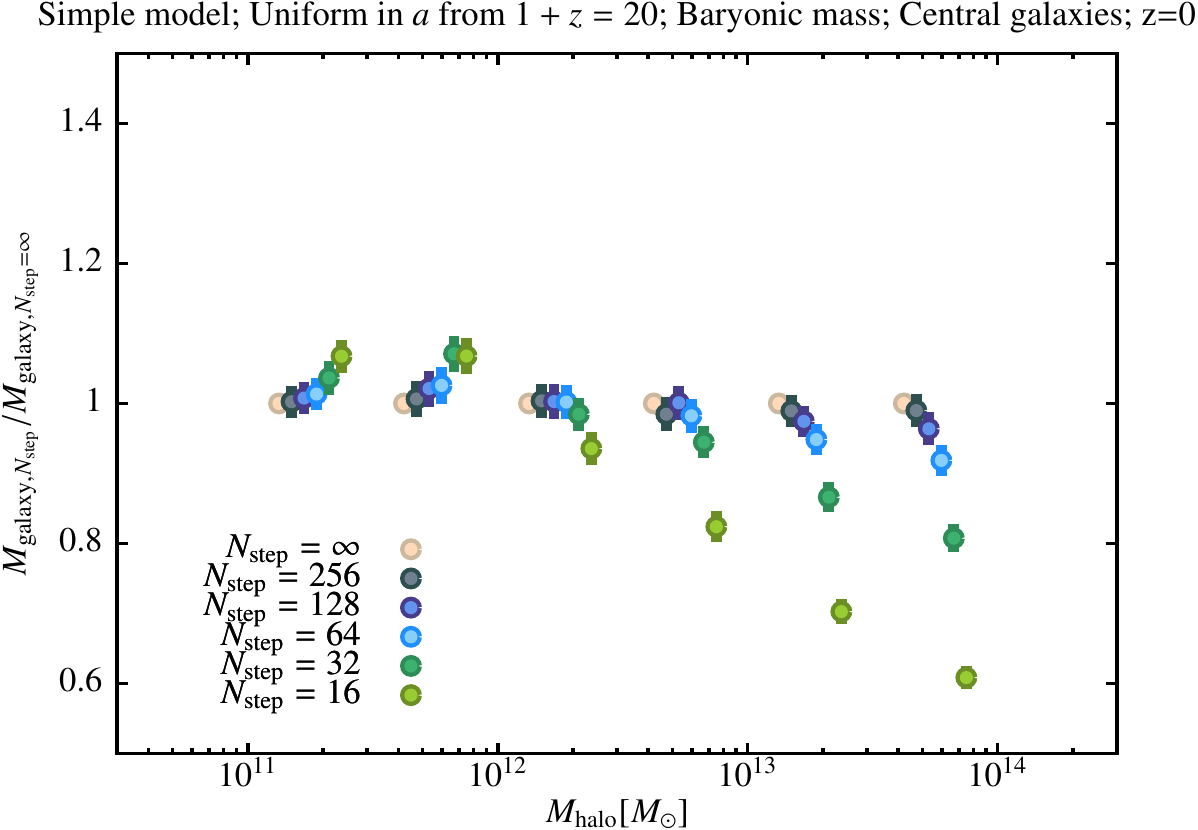} &  \includegraphics[height=60mm]{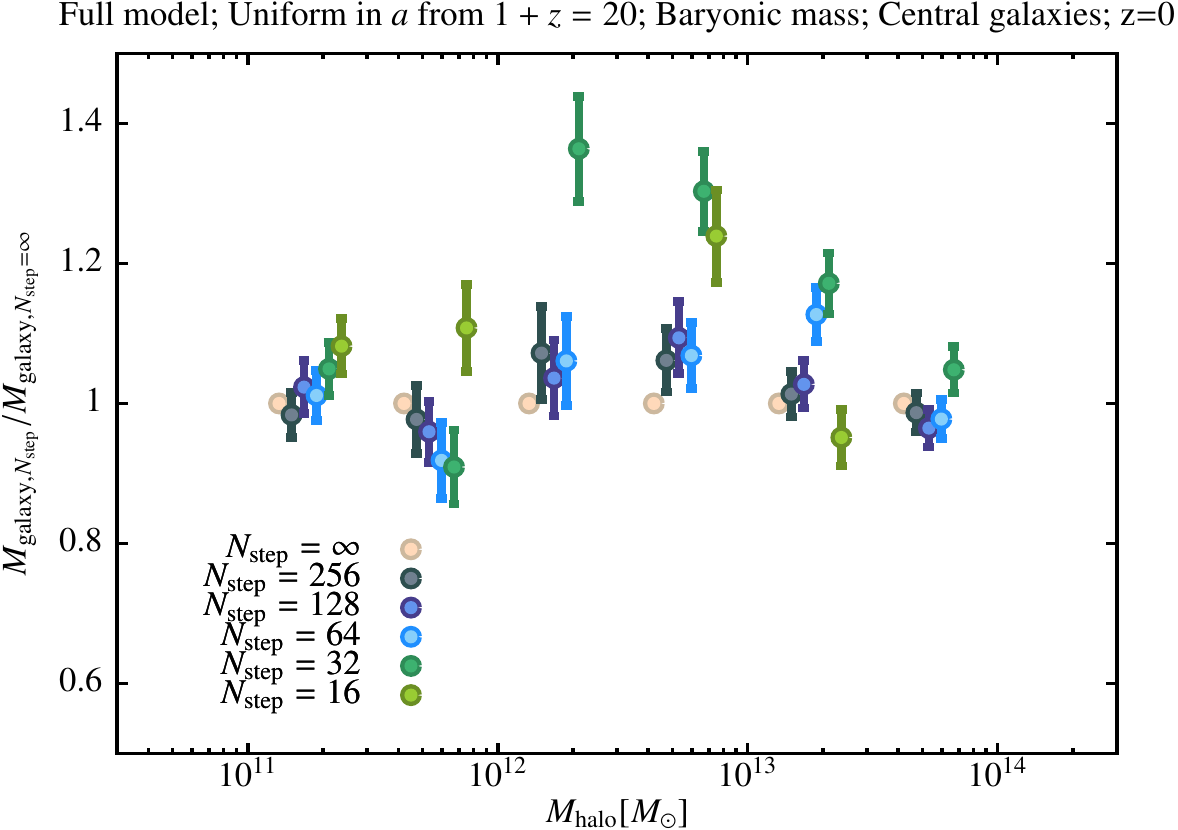} \\
   \includegraphics[height=60mm]{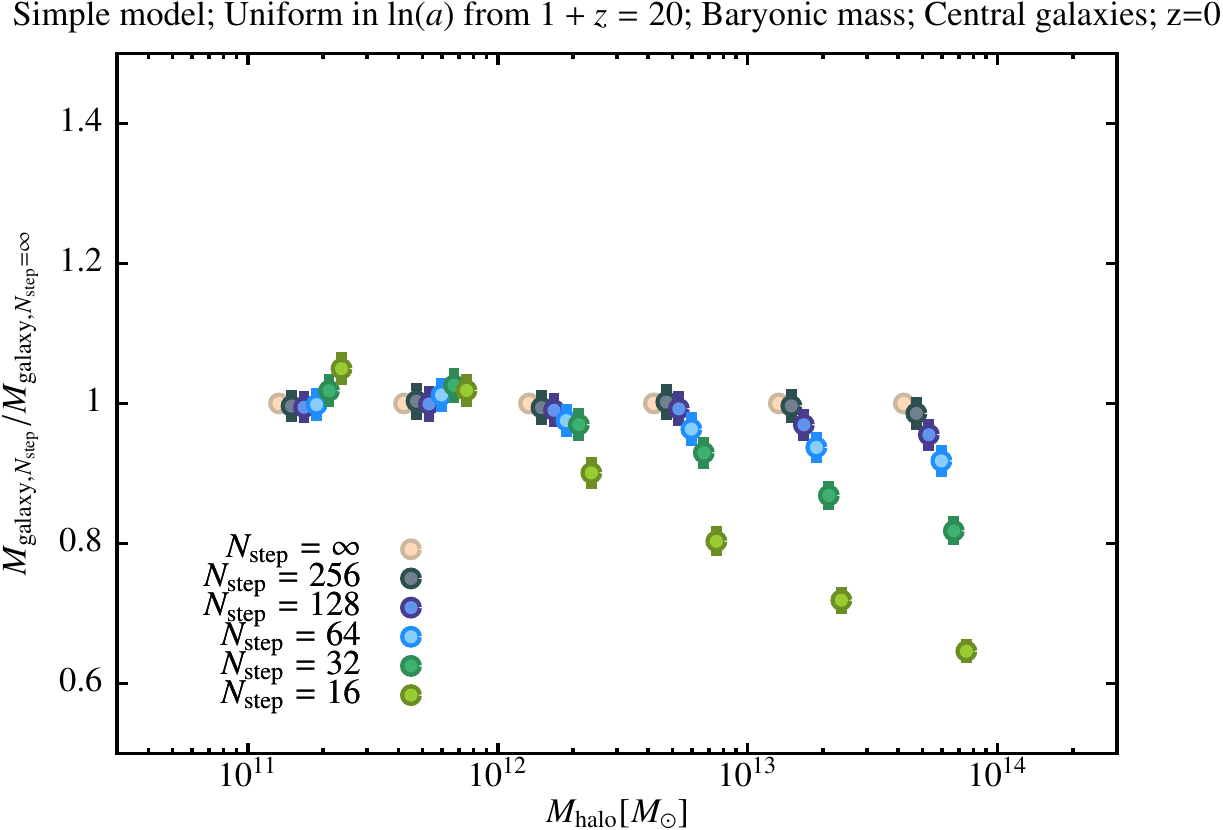} & \includegraphics[height=60mm]{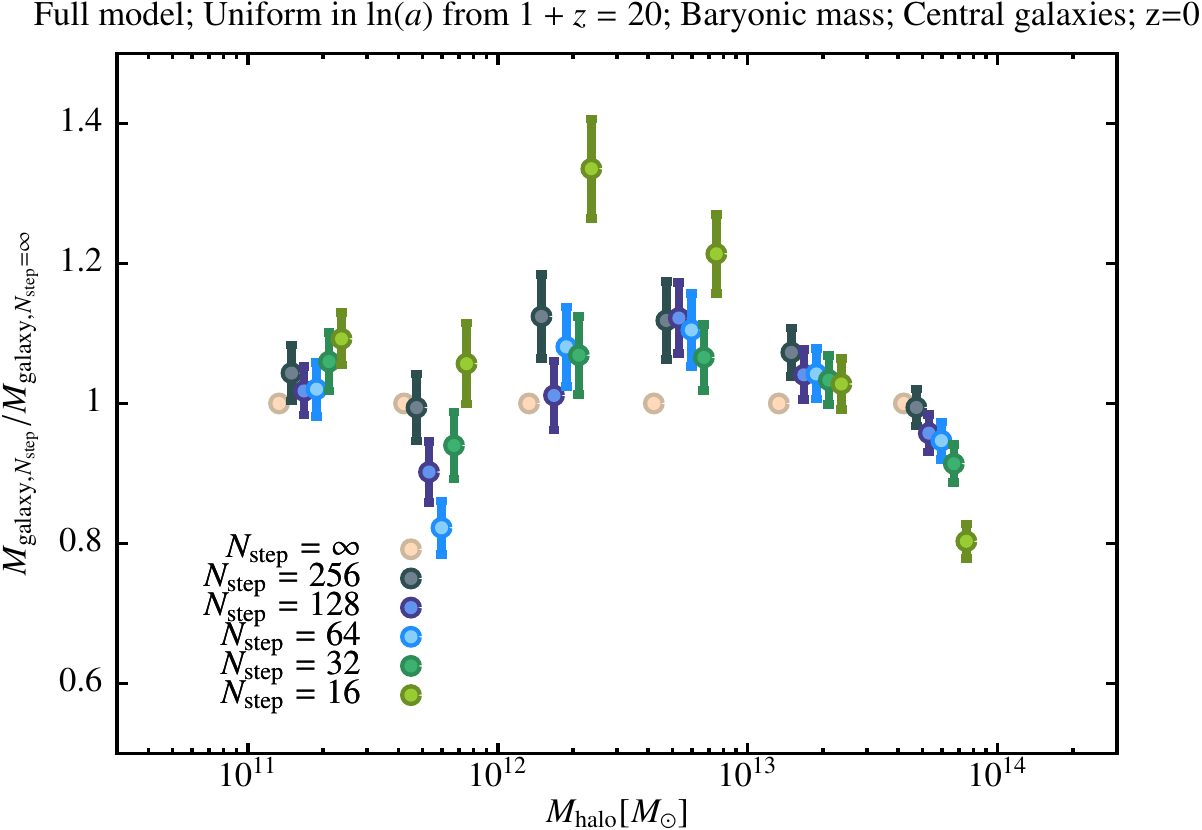} \\
   \includegraphics[height=60mm]{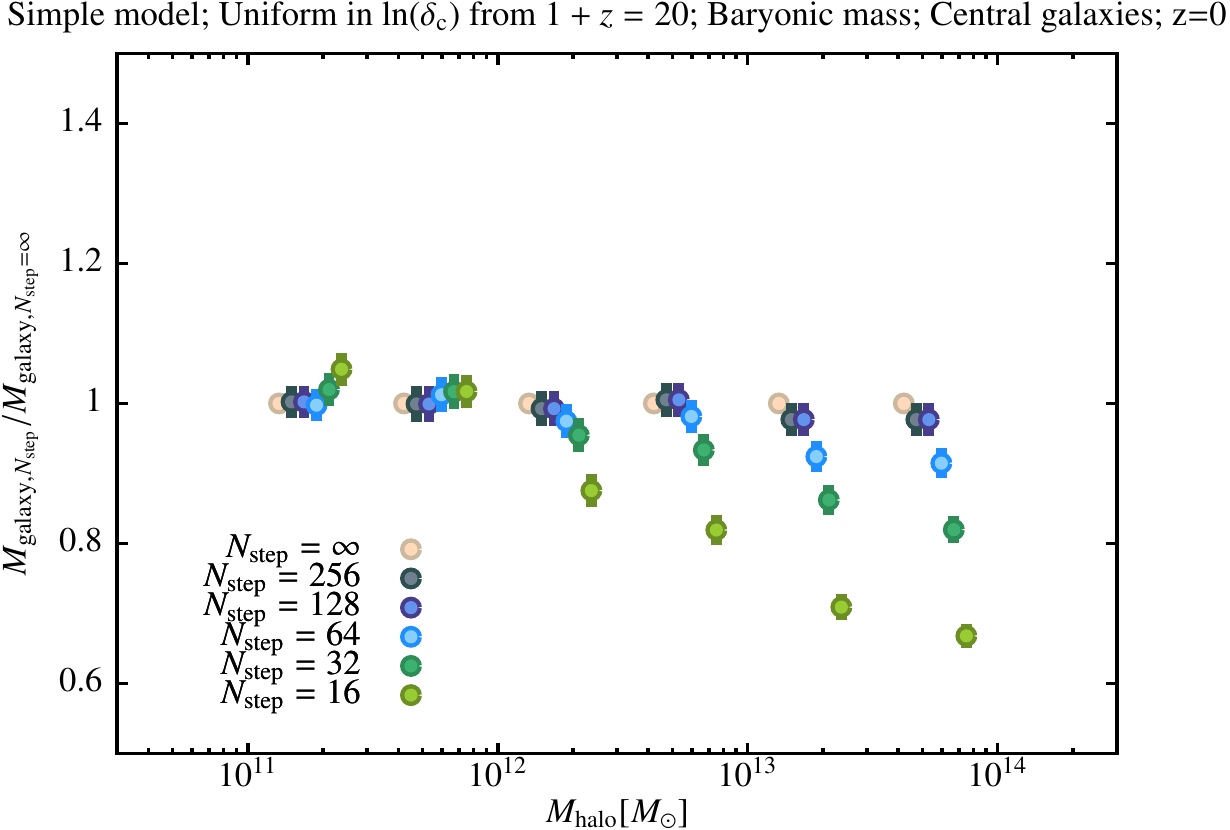} & \includegraphics[height=60mm]{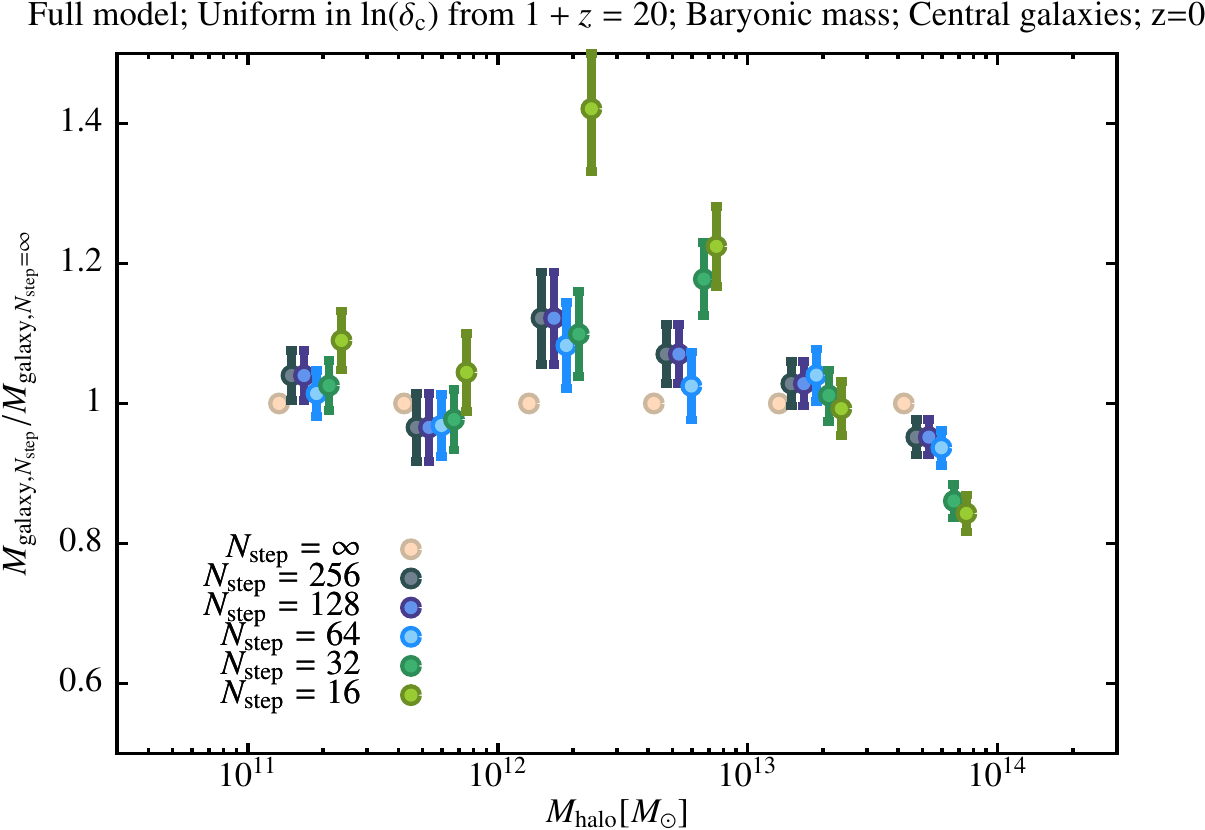} \\
  \end{tabular}
 \end{center}
 \caption{Convergence with number of snapshots of the mean total baryonic mass in central galaxies at $z=0$ as a function of halo mass. Panels in the left column correspond to the simple model, while those in the right column correspond to the full model. Rows correspond to snapshots uniformly spaced in $a$, $\ln(a)$ and $\ln(\delta_{\rm c})$ from top to bottom. In all cases, the earliest snapshot is at $1+z=20$ and the final snapshot at $1+z=1$. Symbol colour corresponds to the number of snapshots used, while error bars indicate the error on the mean mass due to the finite number of merger trees realized in each bin. Points in each mass bin are given small horizontal offsets for clarity.}
 \label{fig:convergenceBaryonicCentrals}
\end{figure*}

Figure~\ref{fig:convergenceBaryonicCentrals} shows results for the convergence in the average total (gaseous plus stellar) mass of central galaxies in bins of dark matter halos mass for both simplified and full models with snapshot spacings uniform in $a$, $\ln a$ and $\ln \delta_{\rm c}$. Symbol colours correspond to different numbers of snapshots as indicated in each panel. In each case we plot the ratio of the mean mass of central galaxies to that found in the $N \rightarrow \infty$ calculations. Error bars on the points are an estimate of the error on the mean due to the finite number of merger trees realized in each mass bin\footnote{In general, this error is largest in the lower mass halos---a consequence of the greater dispersion in galaxy properties in these halos. In massive halos, central galaxies form through multiple mergers which results in an averaging that reduces the scatter in their properties, while when we consider all galaxies in the halo the dispersion is reduced in massive halos simply because we average over many more galaxies.}. For the simple models, it can be seen that in all cases convergence is fastest for lower mass halos, and gets progressively worse for higher mass halos. This could be due to the fact that low mass halos are poorly resolved anyway (due to the finite mass resolution imposed on the trees) and so re-gridding does not lose significant information, whereas high mass halos are well resolved and their merger trees contain substantial information (i.e. they have much ``richer'' formation histories) which is lost by re-gridding\footnote{It is not obvious how this loss of information is best quantified. Information is stored in both the structure of the connected tree and in the labels (mass and time) associated with each node. Simply estimating the information content based on the number of bits required to store a tree is misleading, as the labels associated with nodes are not entirely independent (i.e. along a given branch the mass changes in a smooth and predictable way between bifurcations). One approach is to consider just the connected structure of the tree, ignoring the mass and time information associated with each node. In that case, methods that have been developed for analyzing phylogenetic trees can be adopted. For example, the cladistic information content, CIC, provides a useful measure of the information content of a sparse tree which is assumed to represent an underlying binary tree \protect\citep{thorley_information_1998}. For the trees constructed in this work prior to any re-gridding a $10^{12}M_\odot$ tree has a CIC of approximately 14 bits while a $10^{14}M_\odot$ tree has a CIC of around 5000 bits. Re-gridding these trees onto 32 snapshots spaced uniformly in the logarithm of expansion factor results in an information loss of around 2 and 800 bits respectively.}. Alternatively, it may simply be that the baryonic physics (e.g. cooling) is more sensitive to the details of the merger tree in higher mass systems. We have tested these scenarios by re-running the higher mass merger trees using a mass resolution of $1.2\times 10^{11}M_\odot$ (i.e. a mass 100 times larger than in our standard cases). We find that this results in even slower convergence, suggesting that it is the baryonic physics that matters and not the amount of information in the merger trees themselves. For the full models convergence is worst for intermediate mass halos, becoming more rapid in the highest mass systems.

In all cases, convergence to within 5\% across all masses requires $N_{\rm step}=128$. Convergence occurs at very similar rates irrespective of the choice made for the distribution of snapshots in time. Snapshots distributed uniformly in the logarithm of expansion factor or critical overdensity for collapse result in very marginally faster convergence in some cases, but this seems to be a small effect. Given the (expected) similarity in the results for spacings uniform in $\ln(a)$ and $\ln(\delta_{\rm c})$ we will not show results for spacing uniform in $\ln(\delta_{\rm c})$ in subsequent figures.

At intermediate halo masses, convergence seems to occur from opposite directions in the simple and full models. Specifically, models with low $N_{\rm step}$ systematically under(over)-predict the true mass in the simple(full) model. The effects of supernovae-driven outflows, present in the full model but not the simple model, are ultimately responsible for this difference. In the simple model, low mass dark matter halos are able to efficiently accrete and cool gas into the galaxy phase and then later merge with central galaxies. In low $N_{\rm step}$ trees, many of these halos are missed (they form and are subsumed by larger halos between successive timesteps) and so do not form galaxies. This reduces the mass brought in to central galaxies through merging, causing their mass to be underestimated. In the full model, the supernovae-driven outflows eject most of the mass which condenses into low mass halos, reducing their mass and their subsequent contribution to the growth of central galaxies.

\begin{figure*}
 \begin{center}
  \begin{tabular}{cc}
   \includegraphics[height=60mm]{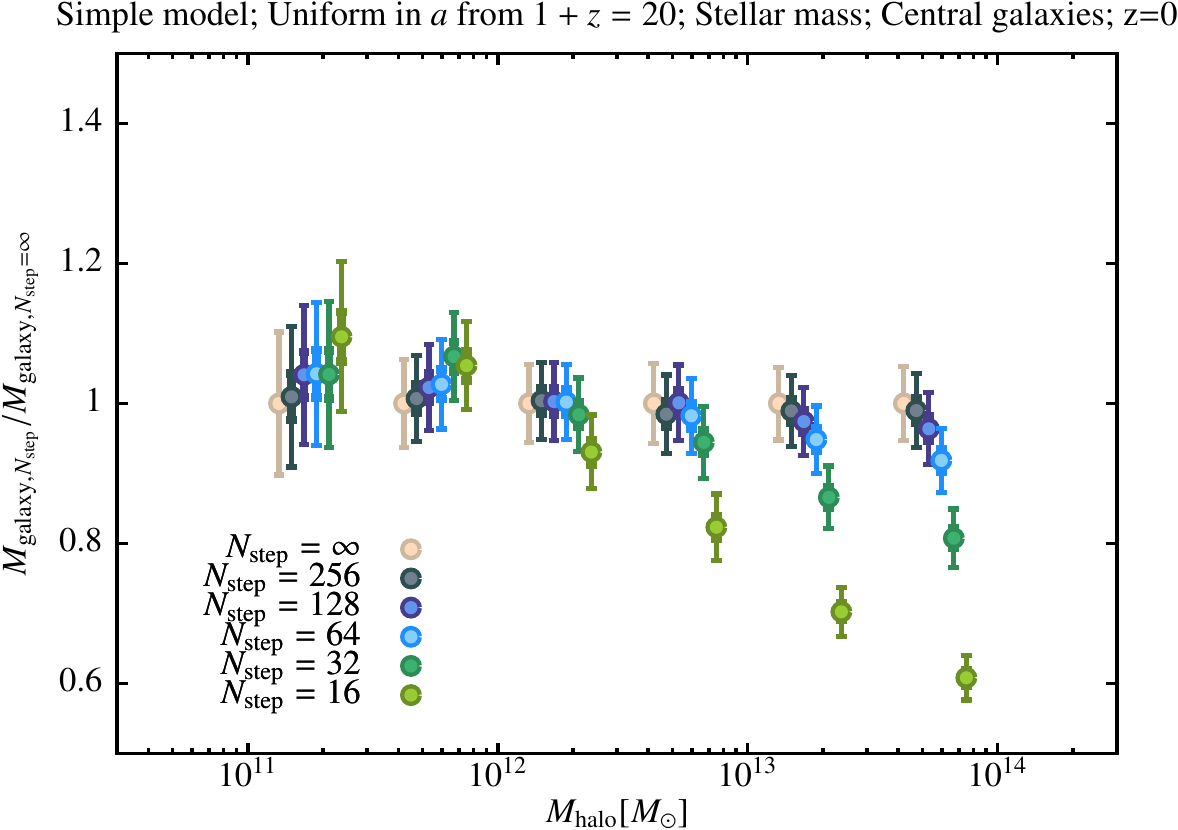} &  \includegraphics[height=60mm]{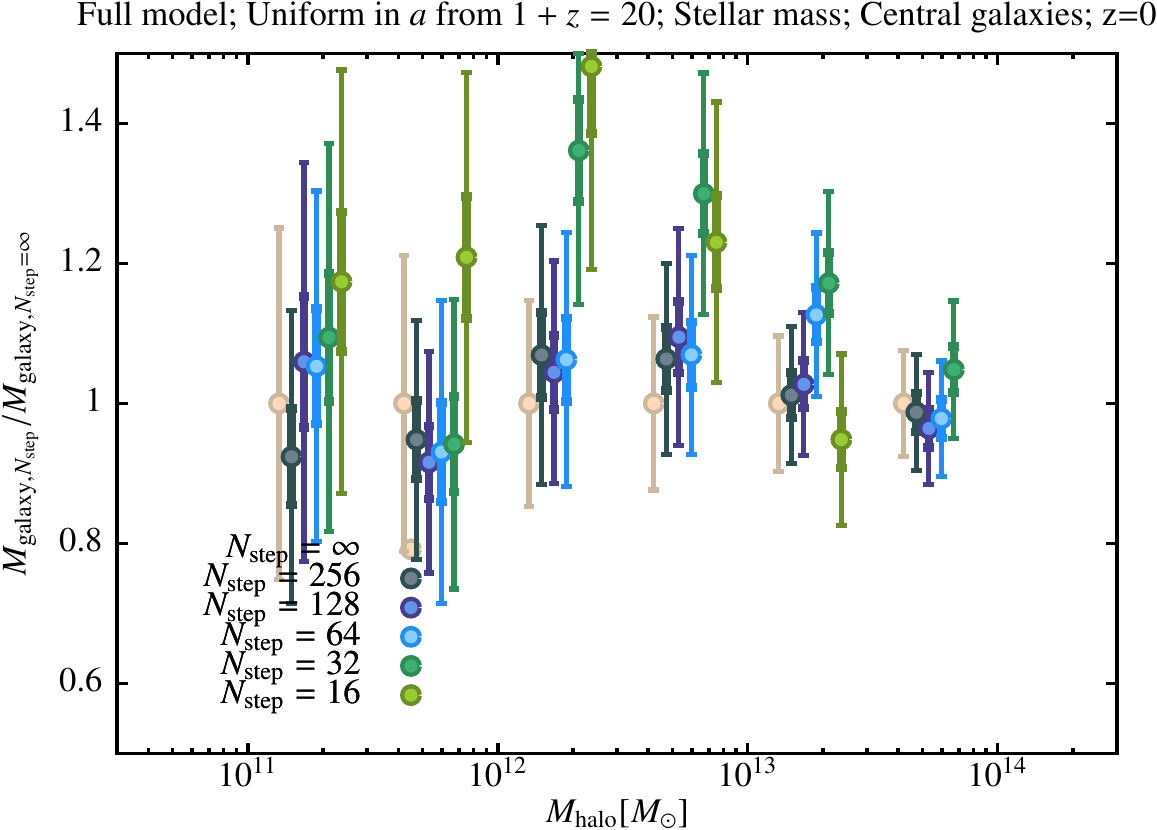} \\
   \includegraphics[height=60mm]{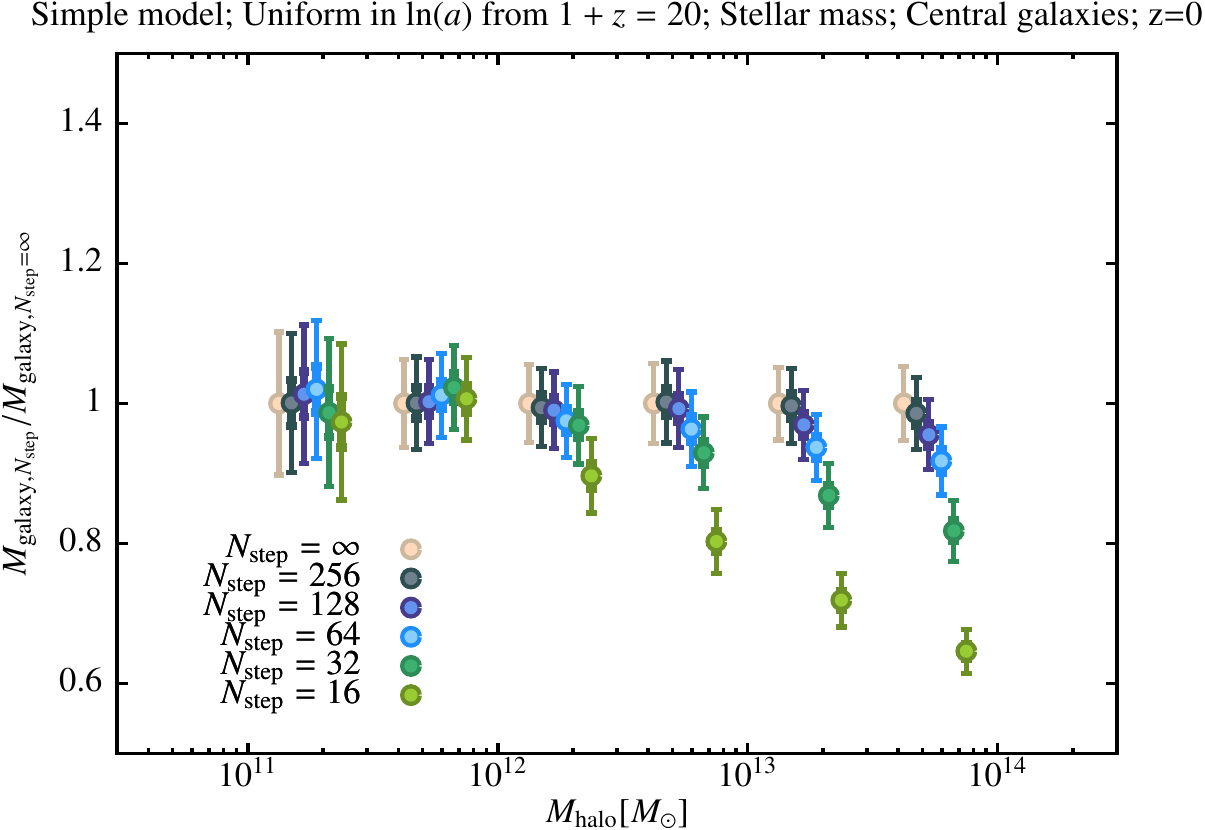} & \includegraphics[height=60mm]{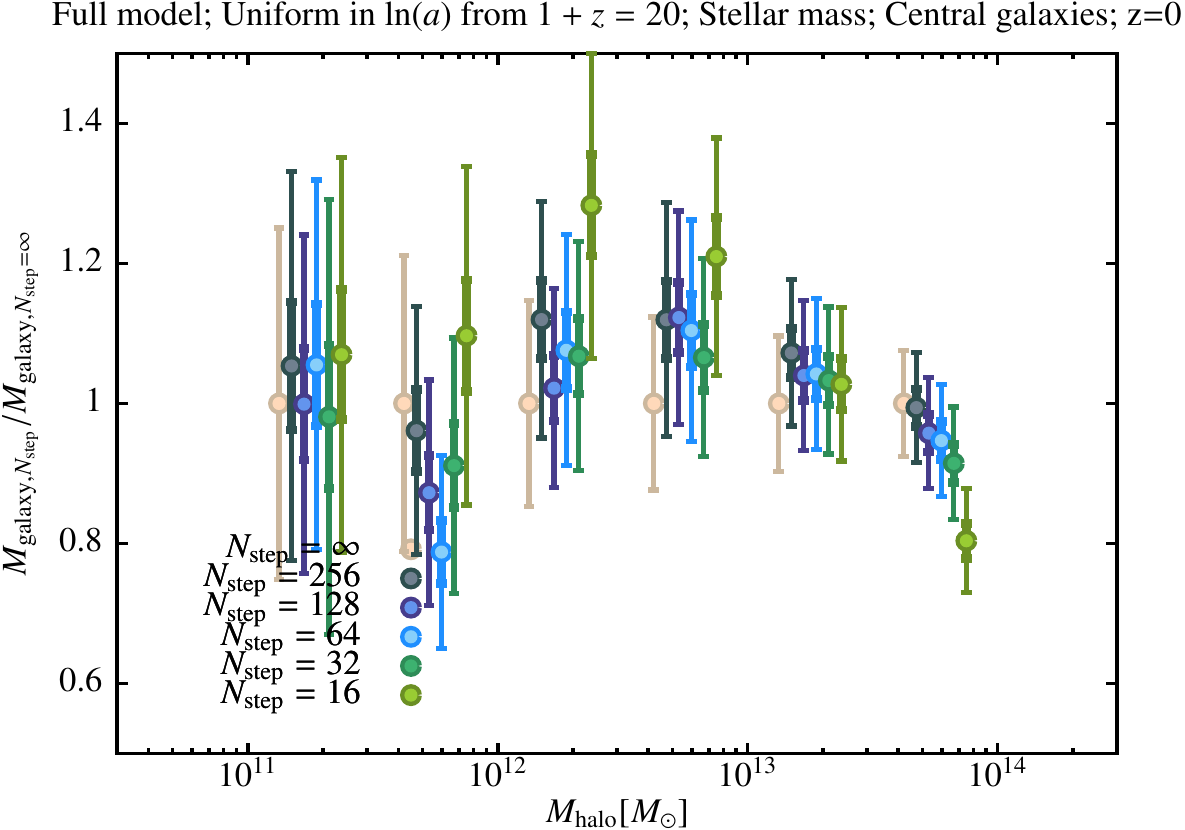} \\
  \end{tabular}
 \end{center}
 \caption{As Fig.~\ref{fig:convergenceBaryonicCentrals} but for the stellar mass of central galaxies. Additionally, we show a second set of errorbars (indicated by thinner lines) which show the root variance (divided by a factor of 5 to keep the error bars smaller than the scale of the y-axis) in the distribution of galaxy stellar masses (as opposed to the error on the mean, shown by the thicker errorbars, which is much smaller).}
 \label{fig:convergenceStellarCentrals}
\end{figure*}

Figure~\ref{fig:convergenceStellarCentrals} shows the same information but now for just the stellar masses of central galaxies at $z=0$. In this figure, we show a second set of errorbars (indicated by thinner lines) which show the root variance (divided by a factor of 5 to keep the error bars smaller than the scale of the y-axis) in the distribution of galaxy stellar masses (as opposed to the error on the mean which is much smaller). These relatively large dispersions in masses illustrate the need for averaging over many merger trees to obtain accurate estimates of the degree of convergence in the mean quantities. The rate of convergence is very similar to the case of total baryonic mass overall, and the same conclusions apply---$N_{\rm step}=128$ is required for convergence to within 5\% across the entire mass range.

\begin{figure*}
 \begin{center}
  \begin{tabular}{cc}
   \includegraphics[height=60mm]{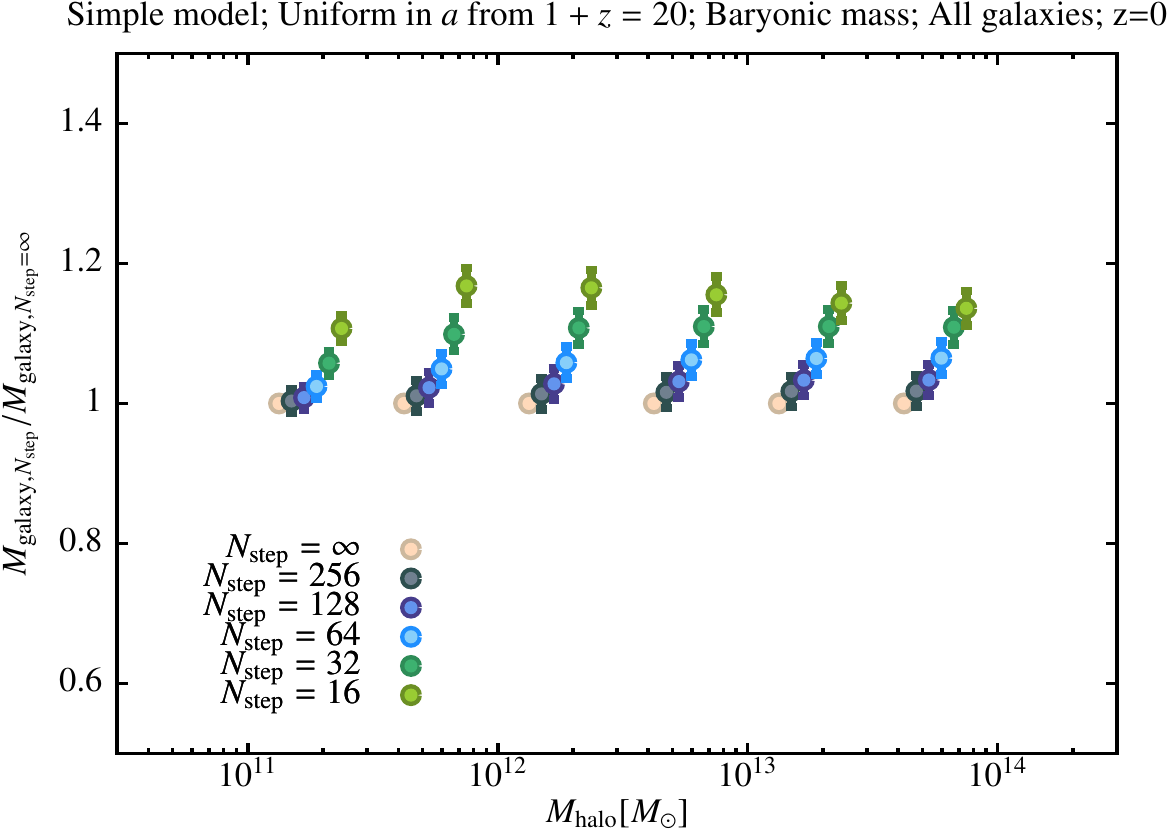} &  \includegraphics[height=60mm]{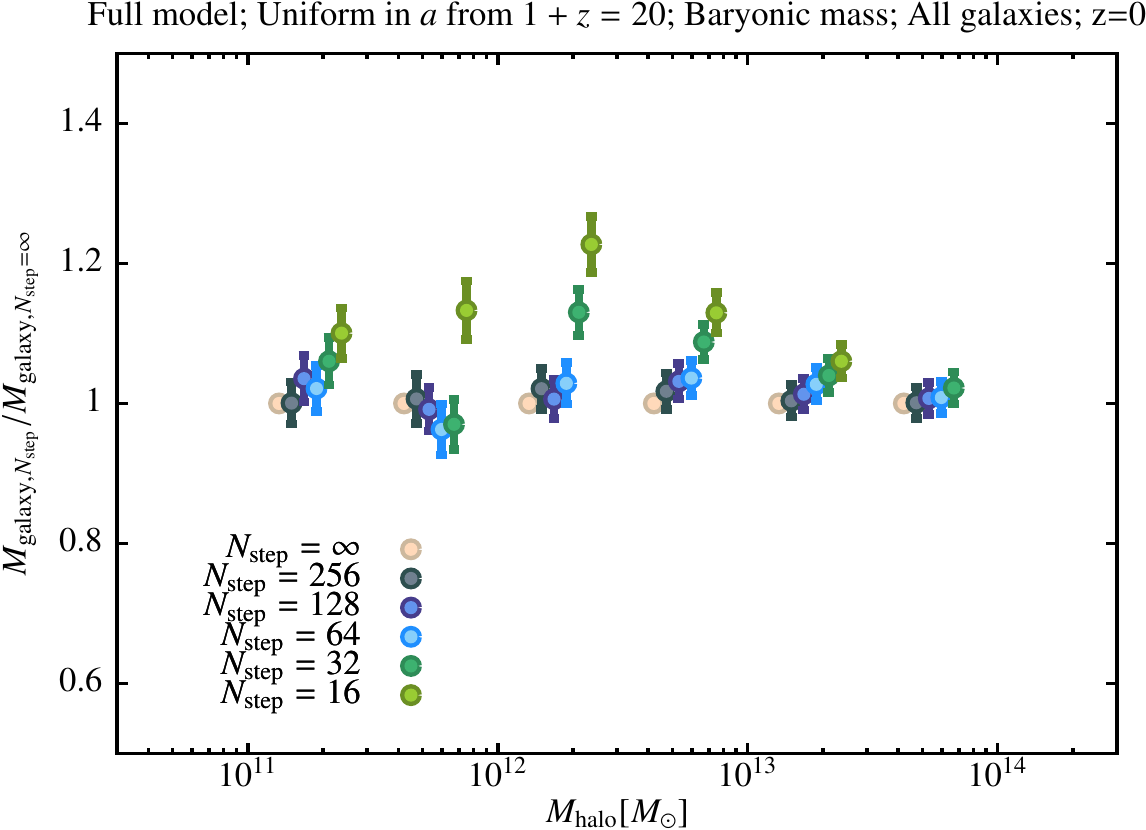} \\
   \includegraphics[height=60mm]{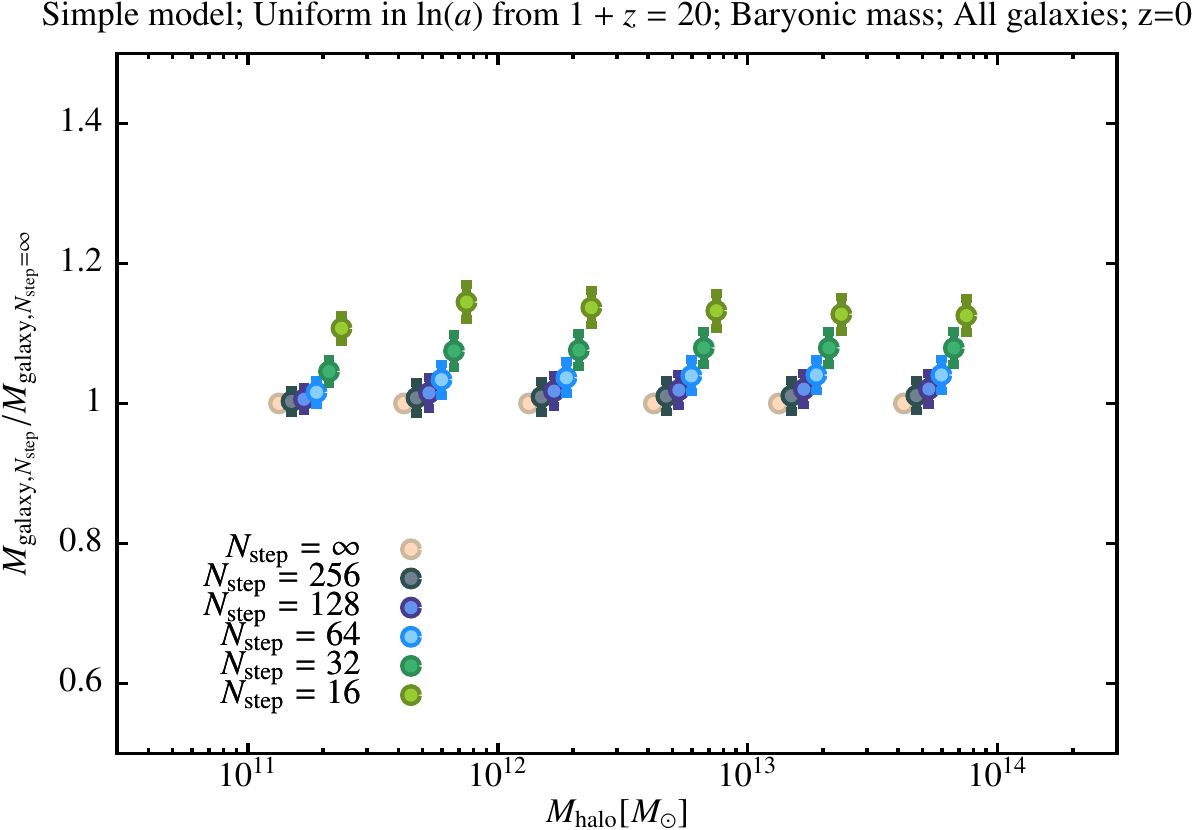} & \includegraphics[height=60mm]{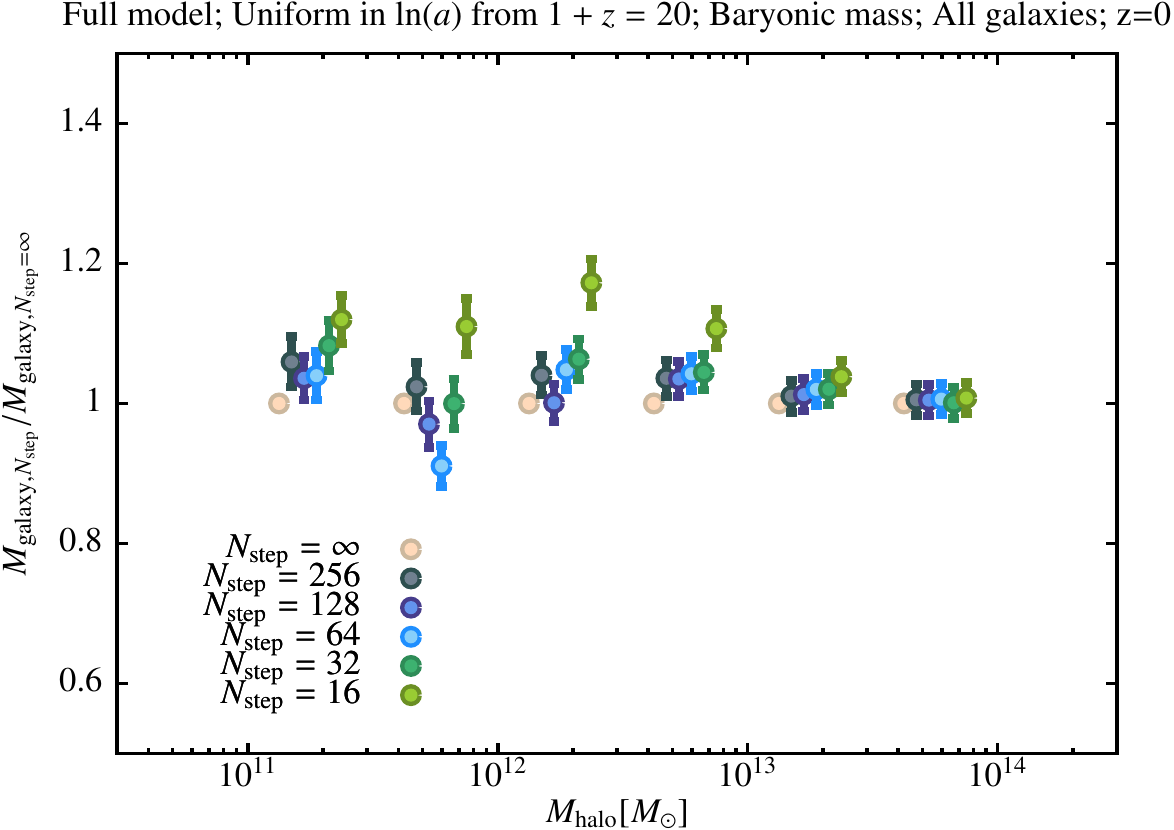} \\
  \end{tabular}
 \end{center}
 \caption{As Fig.~\ref{fig:convergenceBaryonicCentrals} but for the total mass of all galaxies.}
 \label{fig:convergenceBaryonicAlls}
\end{figure*}

\begin{figure*}
 \begin{center}
  \begin{tabular}{cc}
   \includegraphics[height=60mm]{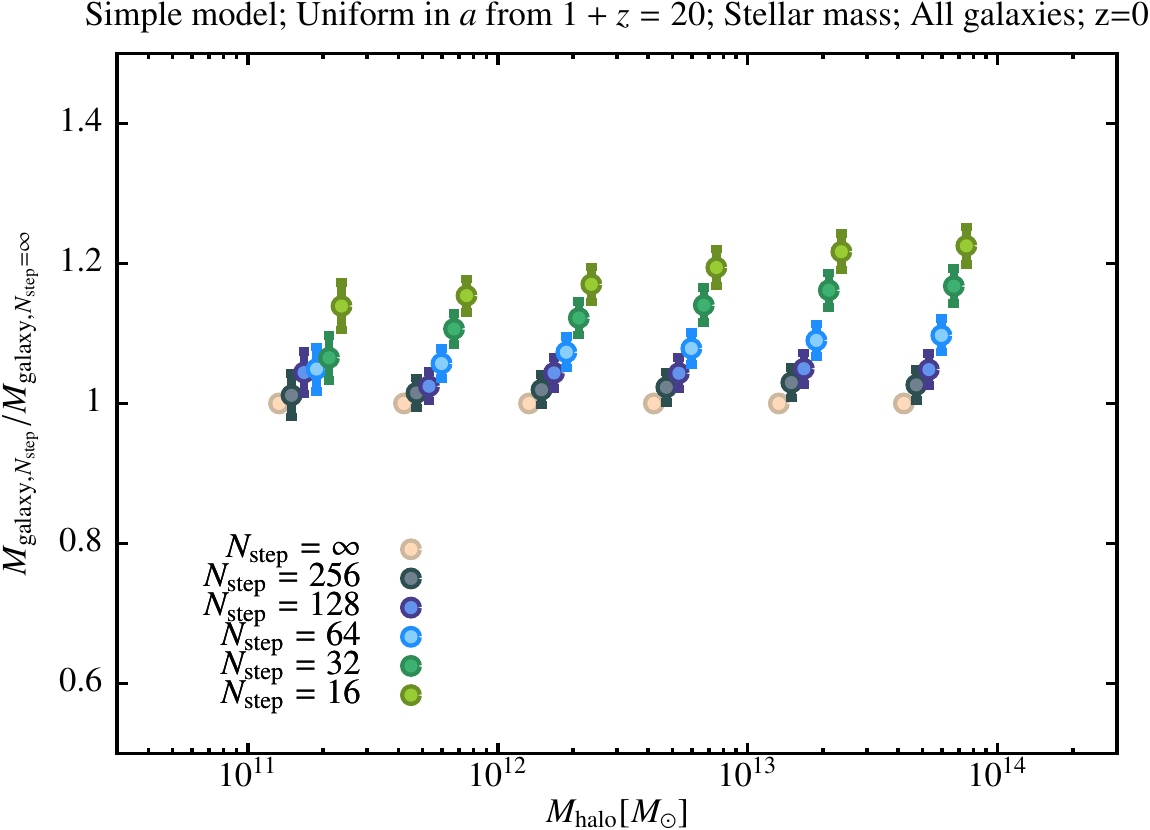} &  \includegraphics[height=60mm]{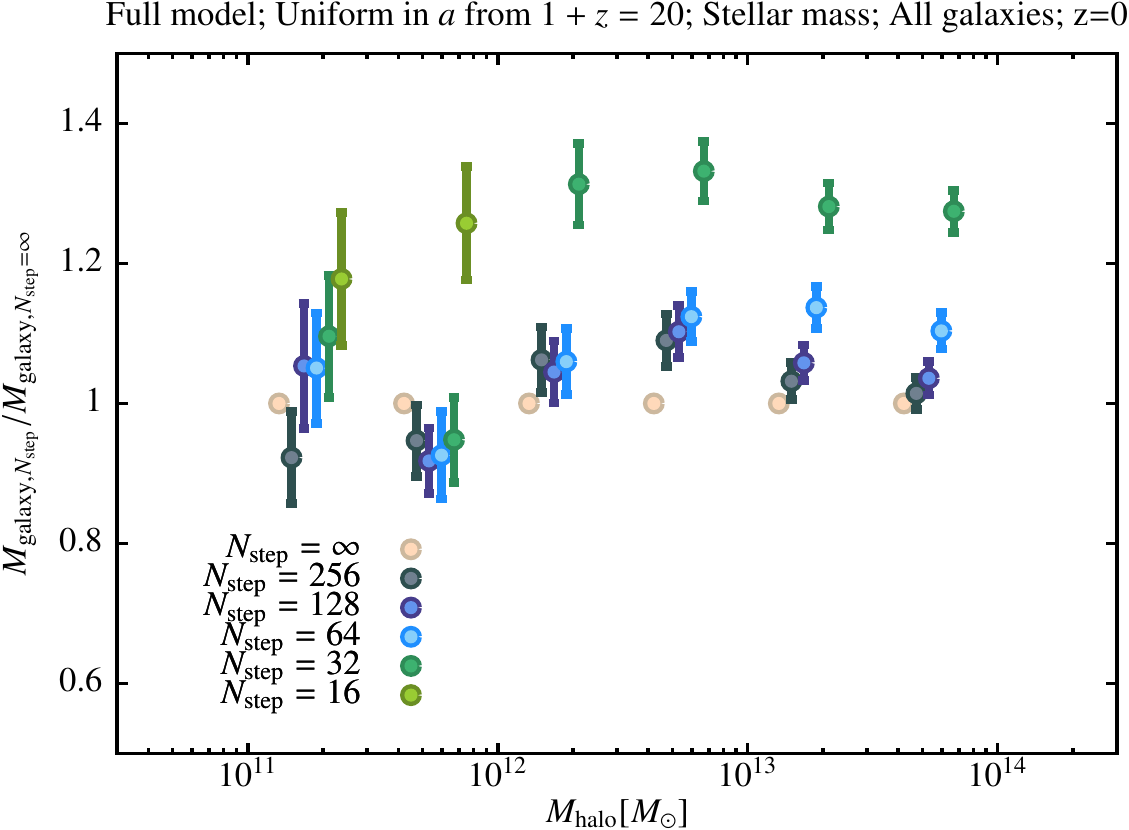} \\
   \includegraphics[height=60mm]{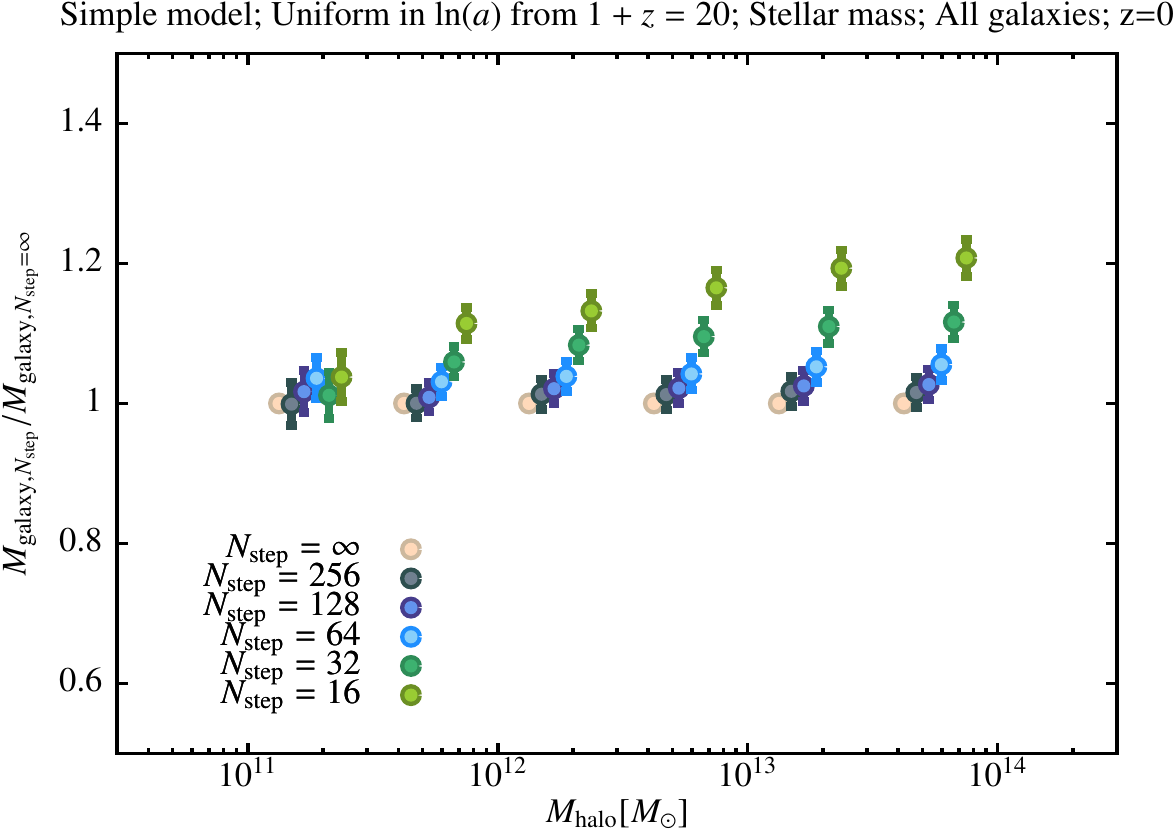} & \includegraphics[height=60mm]{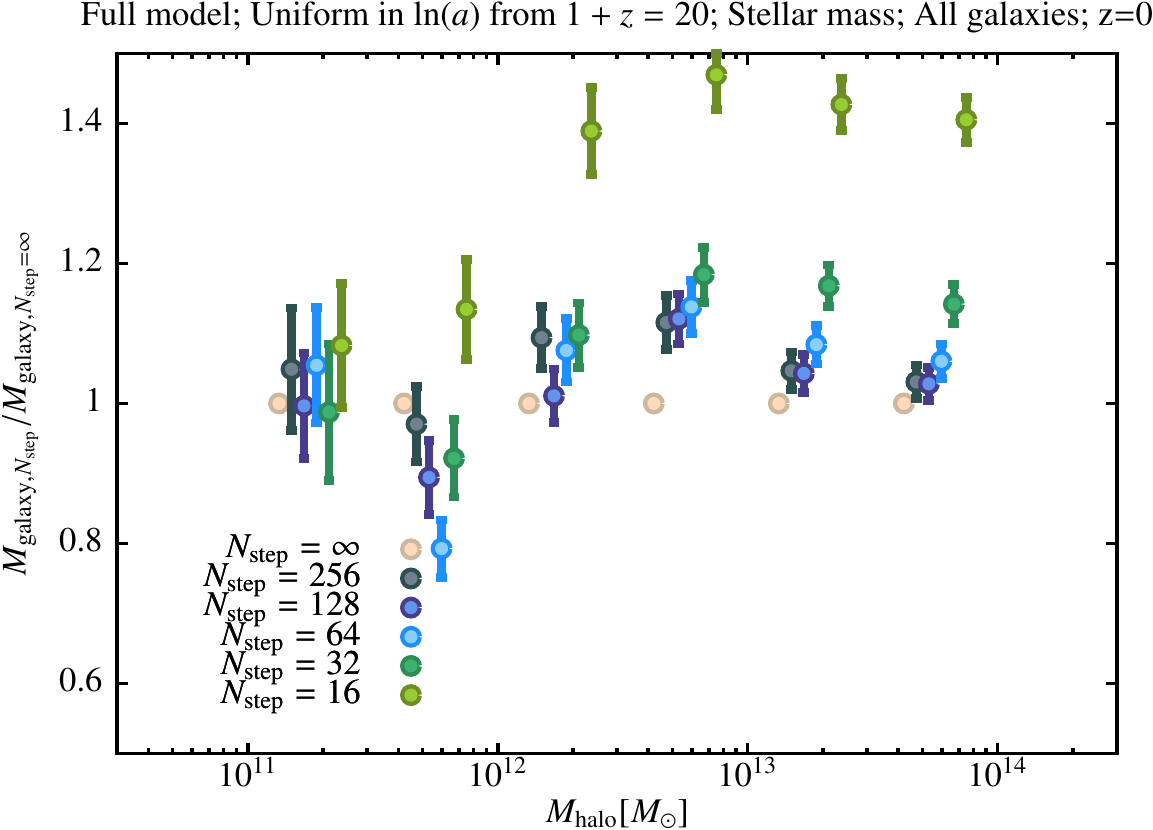} \\
  \end{tabular}
 \end{center}
 \caption{As Fig.~\ref{fig:convergenceBaryonicCentrals} but for the stellar mass of all galaxies.}
 \label{fig:convergenceStellarAlls}
\end{figure*}

When we consider all galaxies (i.e. we sum the masses of all galaxies, satellites and centrals, in a halo and then find the mean of this quantity over many realizations) results change somewhat as shown in Fig.~\ref{fig:convergenceBaryonicAlls} and \ref{fig:convergenceStellarAlls}. For example, the systematic offset when $N_{\rm step}$ is small is in the opposite direction for high mass halos, now always over-predicting the mass. We find that $N_{\rm step}=64$ is sufficient for 5\% convergence in total baryonic mass, while $N=128$ is required for the same degree of convergence in stellar mass. 

\begin{figure*}
 \begin{center}
  \begin{tabular}{cc}
   \includegraphics[height=60mm]{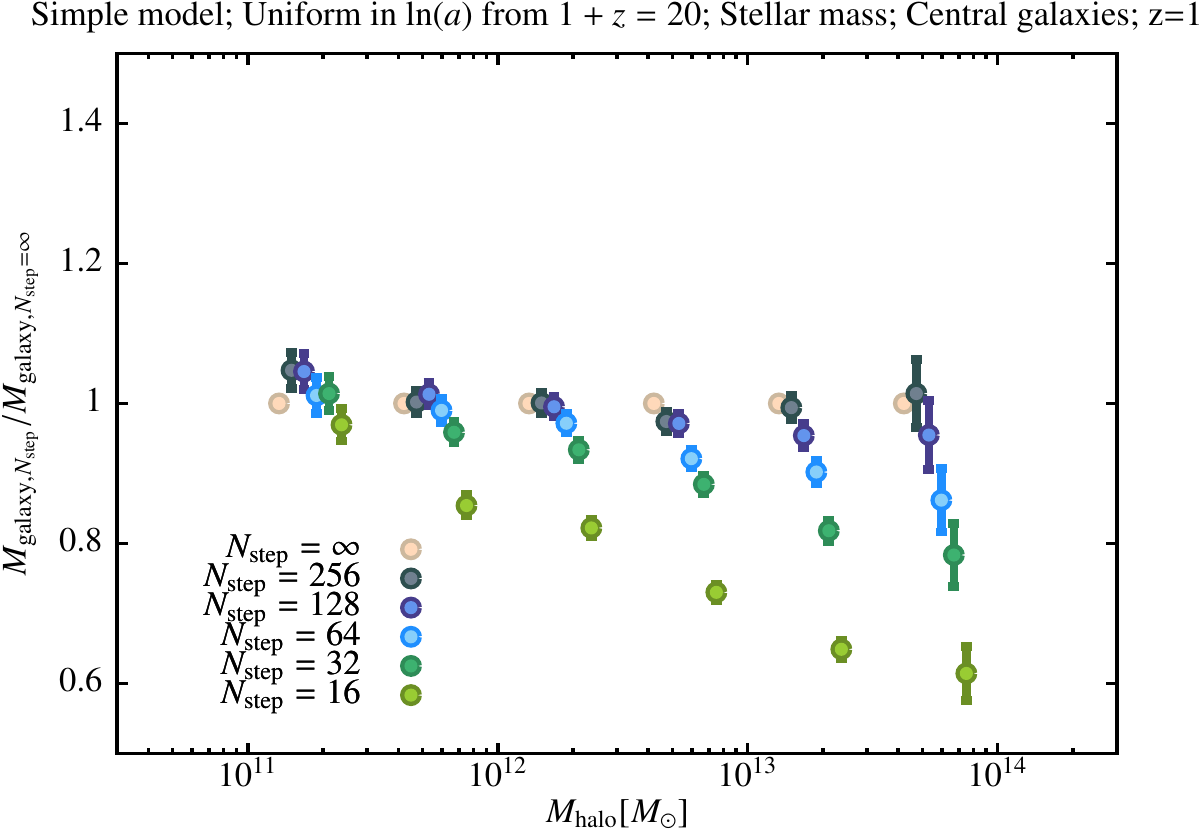} &  \includegraphics[height=60mm]{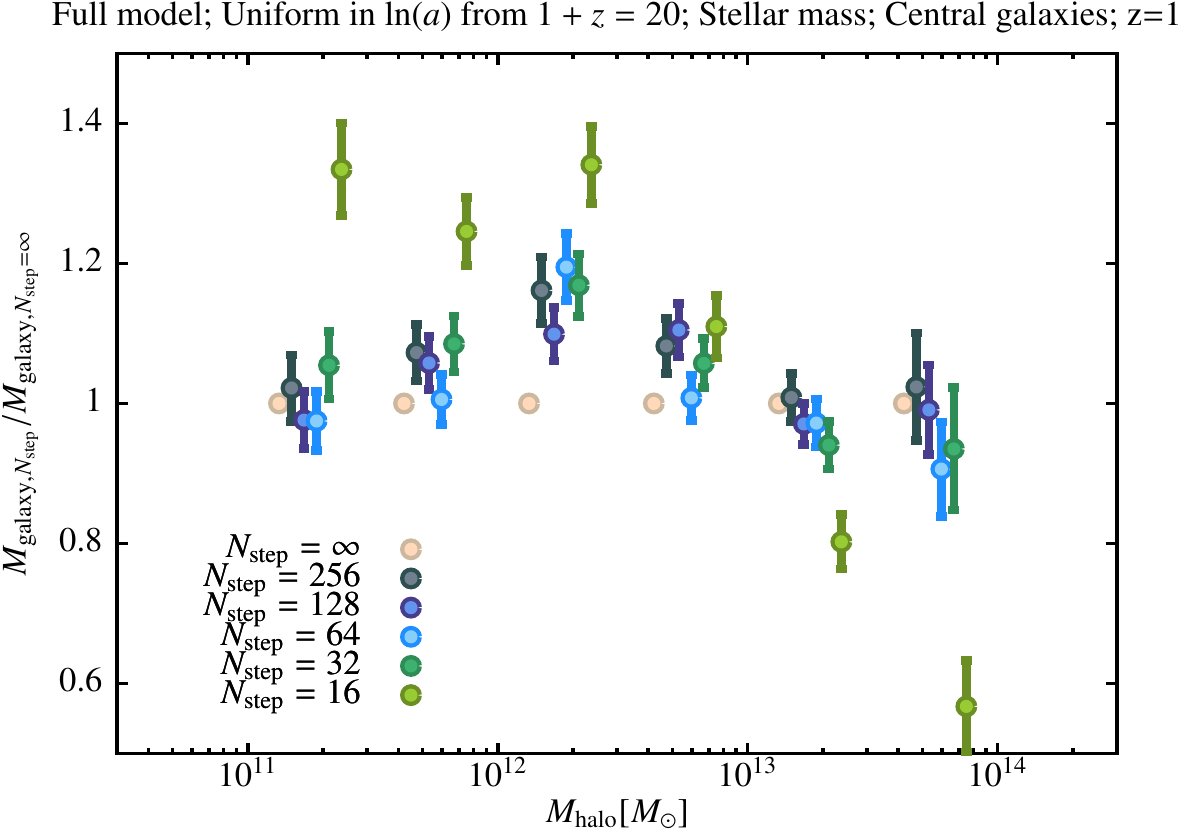} \\
   \includegraphics[height=60mm]{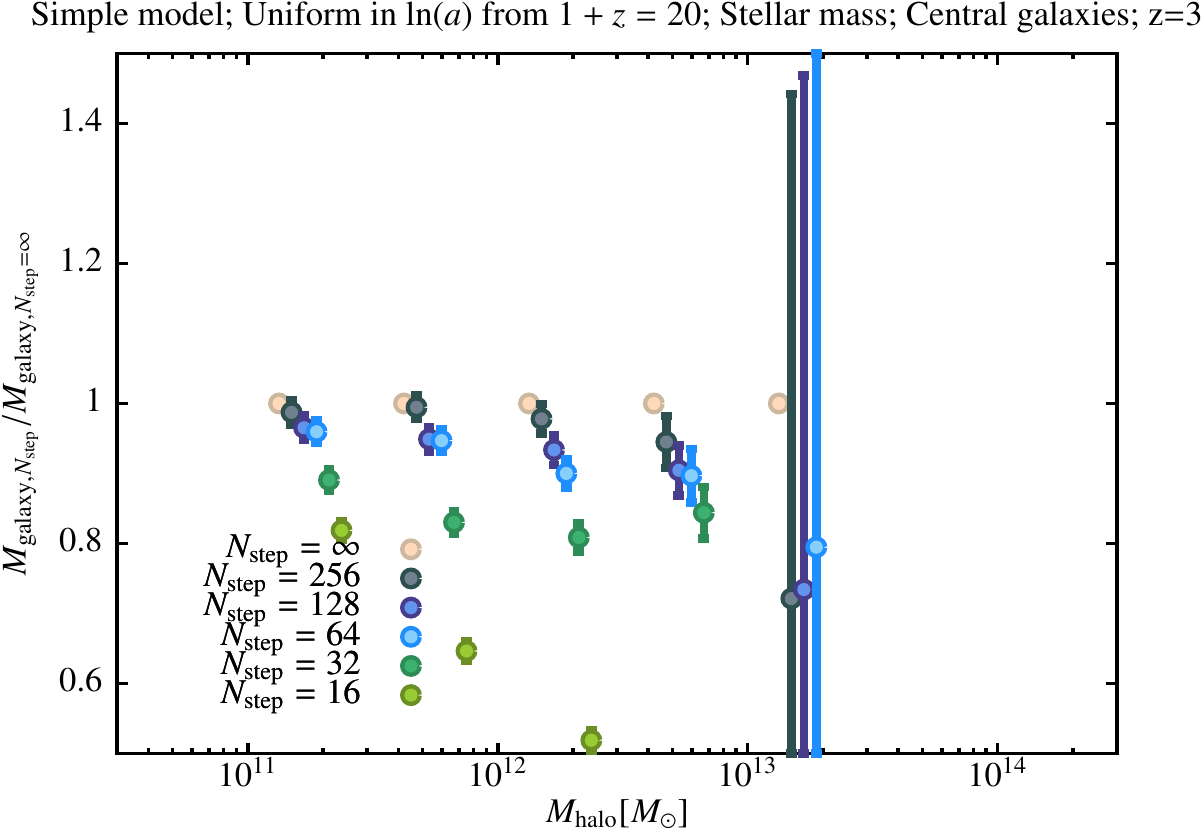} & \includegraphics[height=60mm]{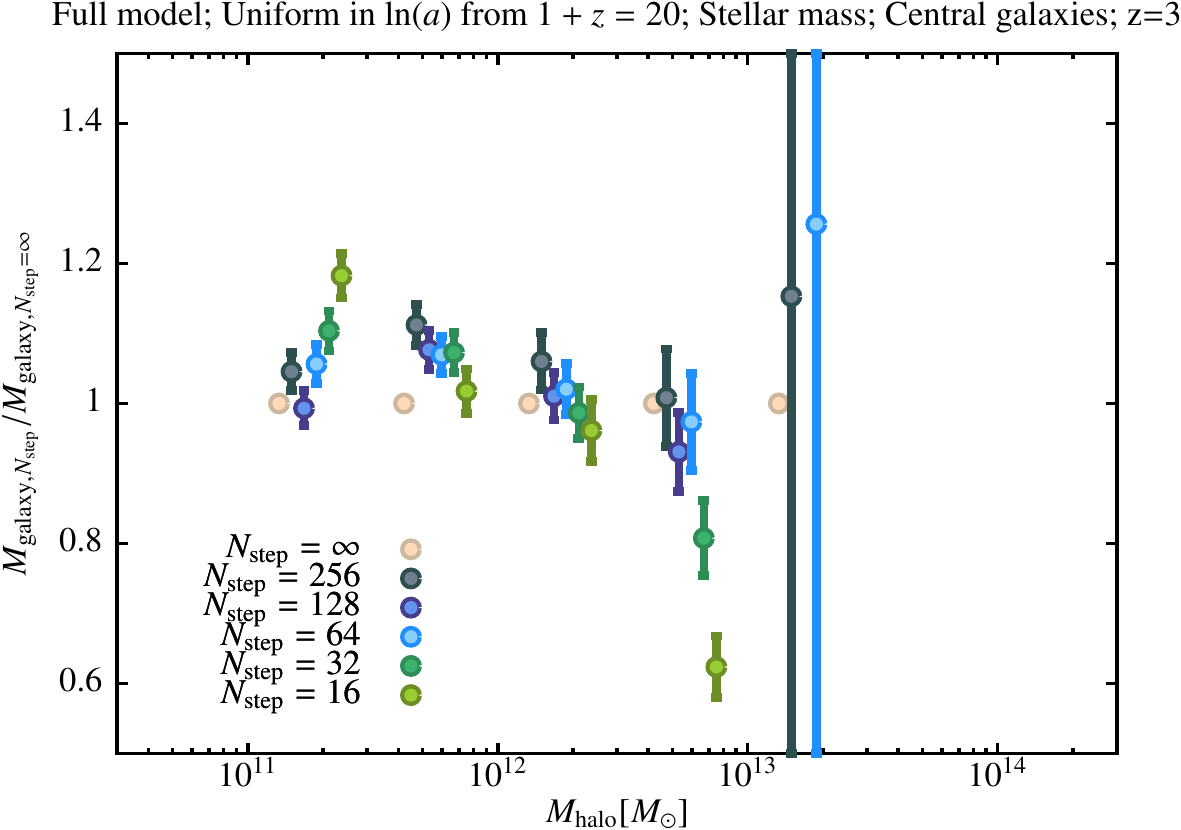} \\
  \end{tabular}
 \end{center}
 \caption{As Fig.~\ref{fig:convergenceBaryonicCentrals} but for the stellar mass of central galaxies at $z>0$. All results use snapshots uniformly spaced in $\ln(a)$. Rows indicate results for $z=1$ and 3 (from top to bottom).}
 \label{fig:convergenceStellarHiZ}
\end{figure*}

Figure~\ref{fig:convergenceStellarHiZ} shows convergence in the stellar mass of central galaxies at $z=1$ and 3, as a function of their halo mass at those redshifts. Note that the number of snapshots always refers to the total number from $z=20$ to $z=0$ even when results are shown for $z>0$. Errors grow rapidly with increasing redshift. Convergence is slower than for $z=0$, particular in the case of low mass halos at $z=3$. The properties of galaxies at these redshifts depend only on the structure of their progenitor tree at yet higher redshifts, such that the number of relevant snapshots is significantly less than $N_{\rm step}$. 

\subsubsection{Numbers of Galaxies}

\begin{figure*}
 \begin{center}
  \begin{tabular}{cc}
   \includegraphics[height=60mm]{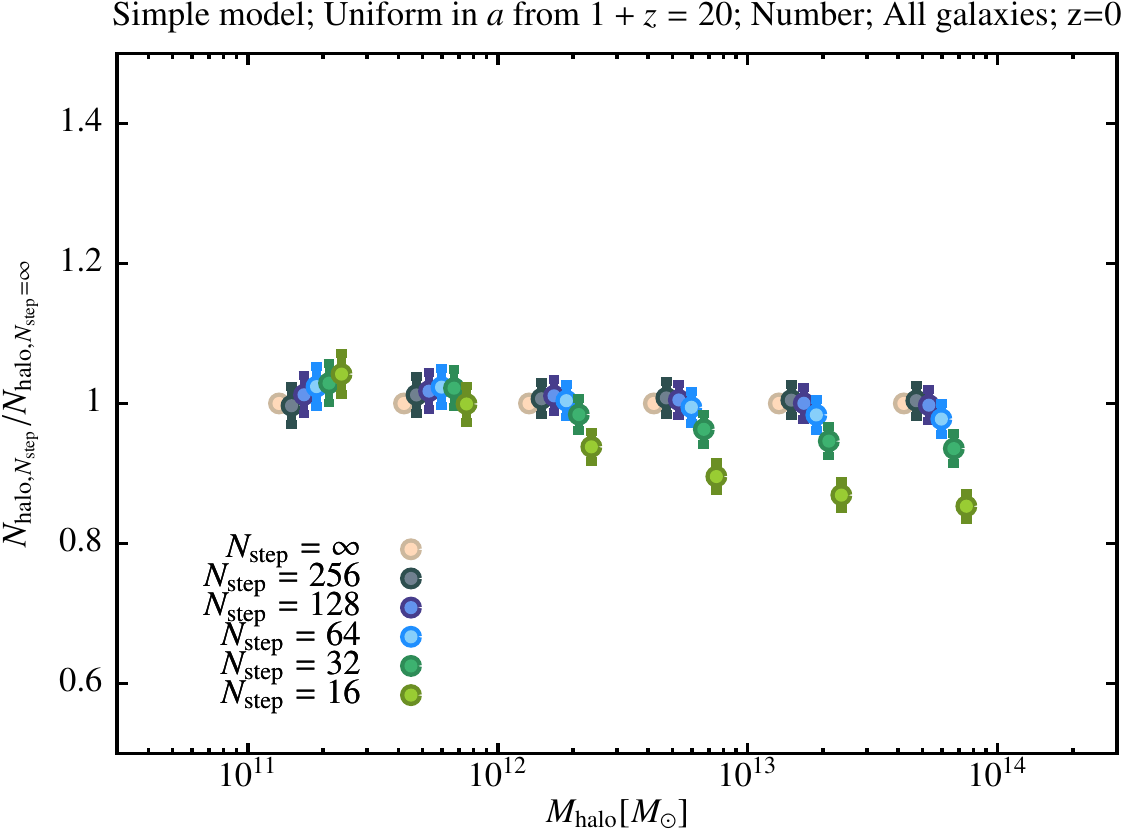} &  \includegraphics[height=60mm]{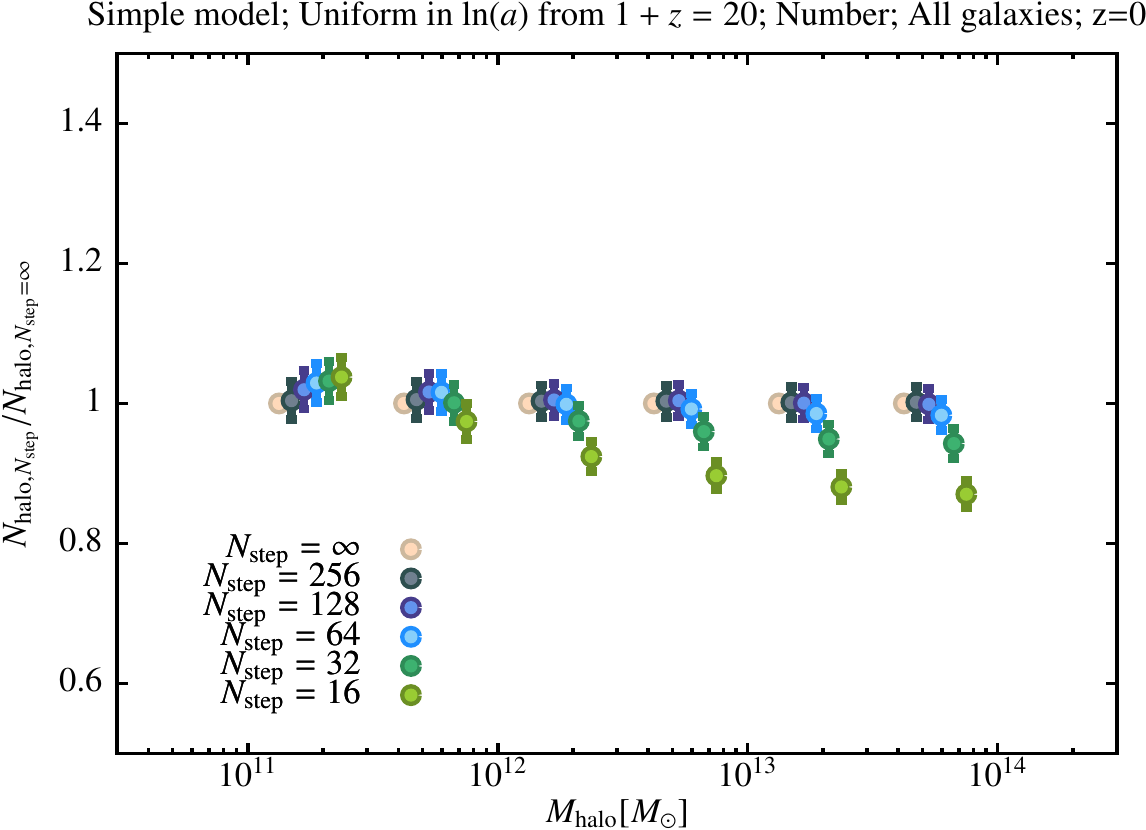} \\
   \includegraphics[height=60mm]{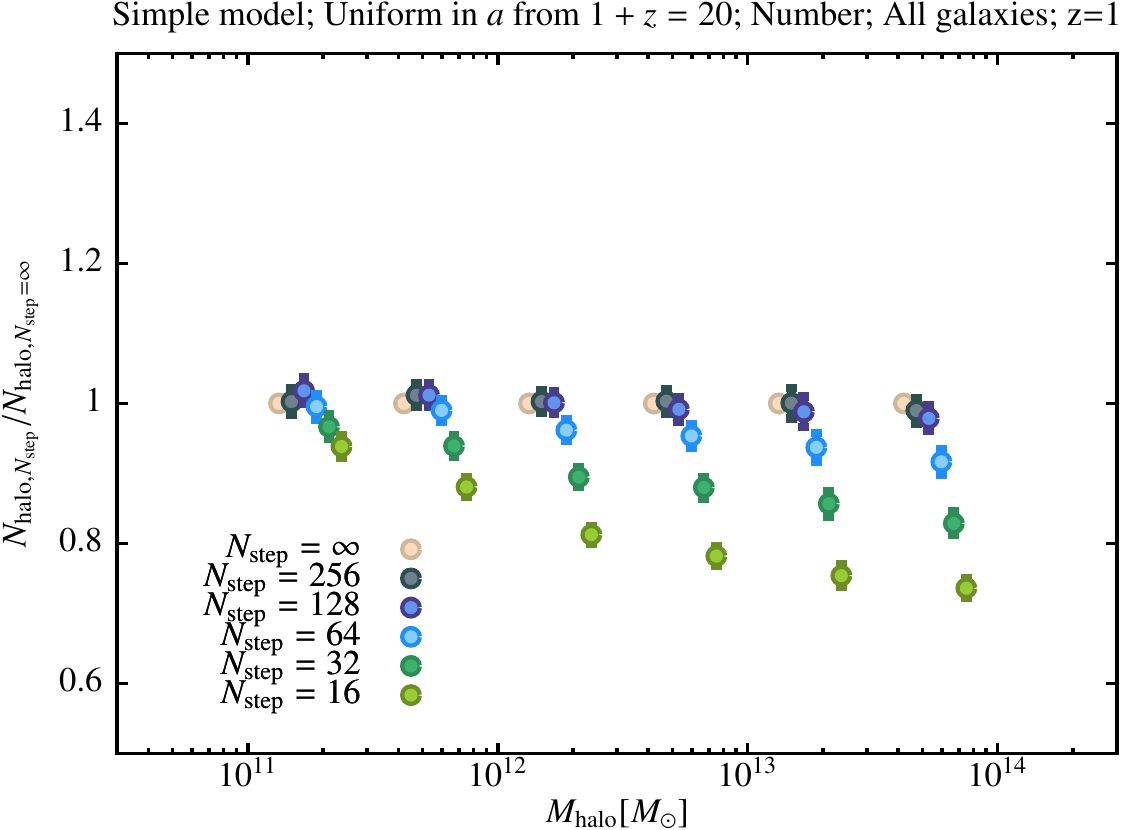} &  \includegraphics[height=60mm]{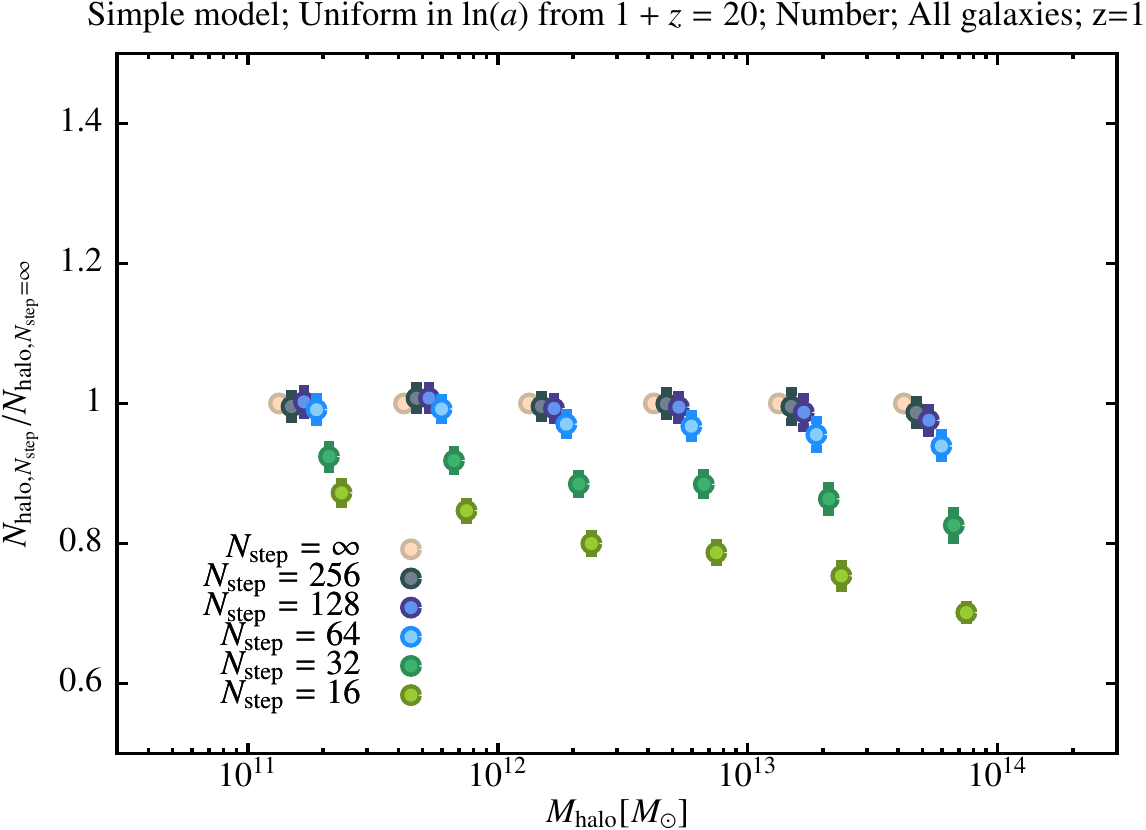} \\
   \includegraphics[height=60mm]{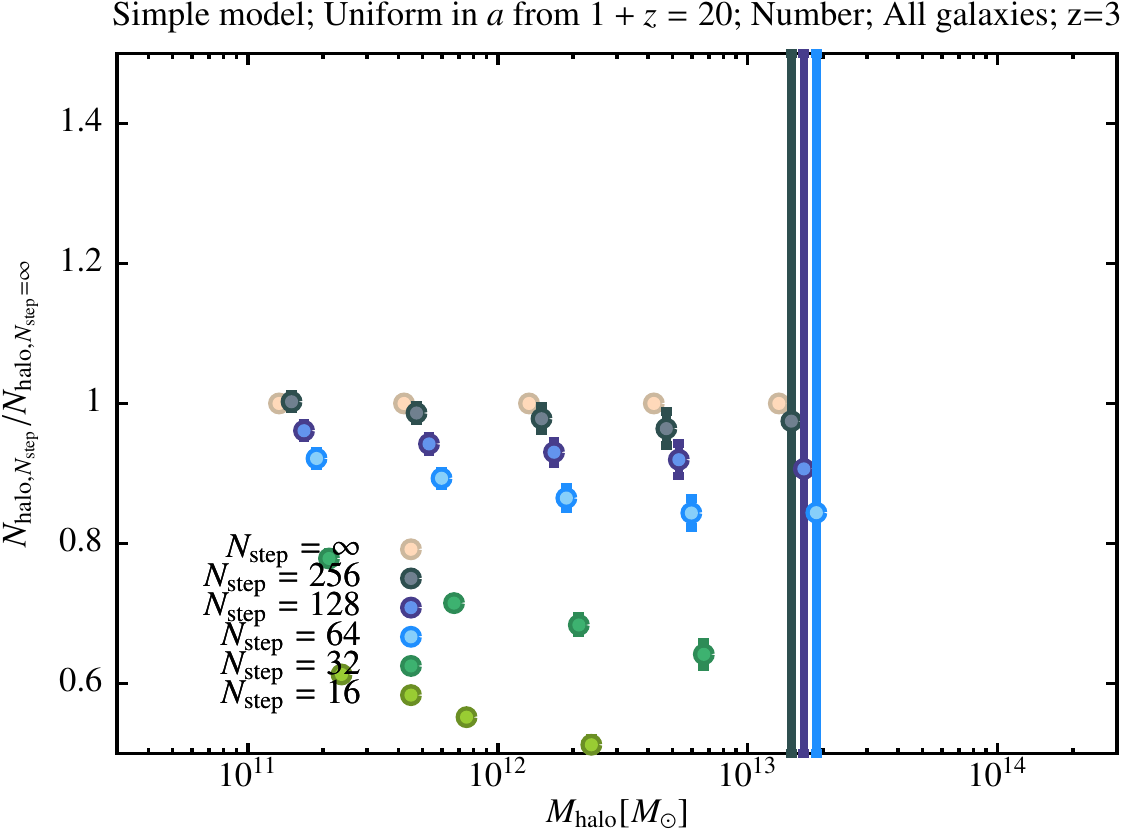} &  \includegraphics[height=60mm]{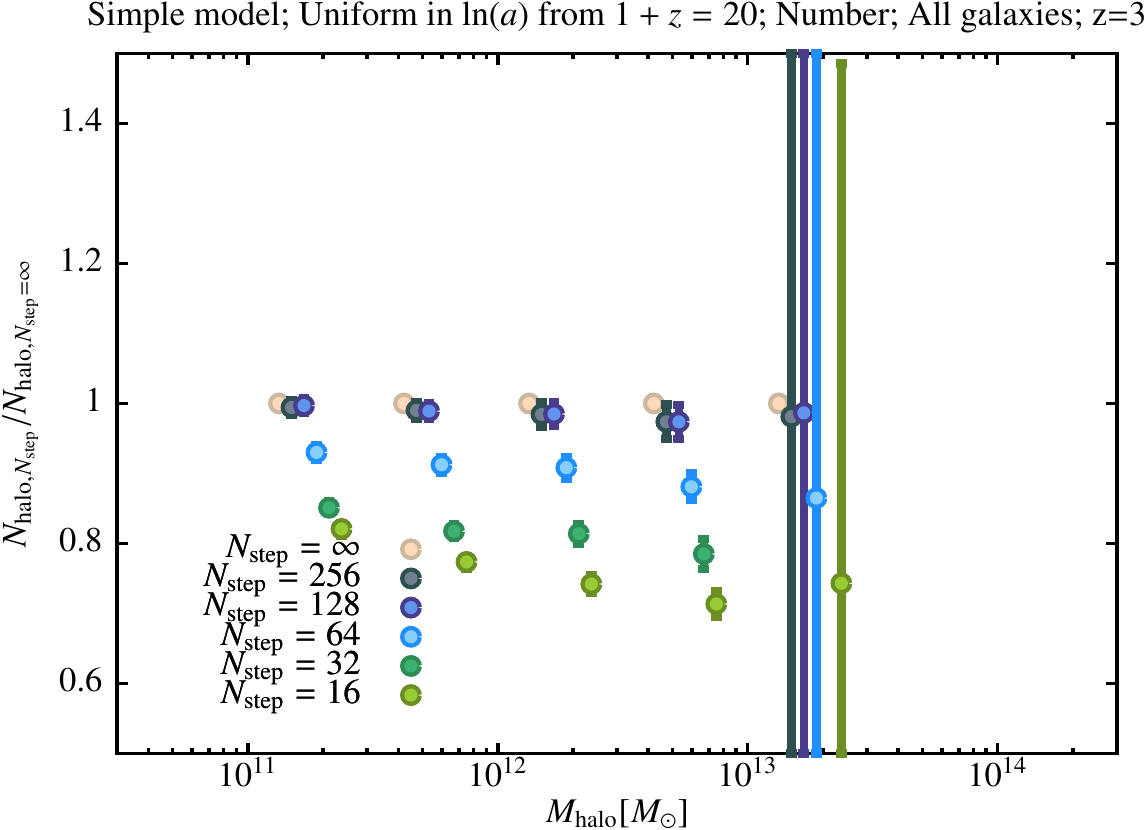} \\
  \end{tabular}
 \end{center}
 \caption{As Fig.~\ref{fig:convergenceBaryonicCentrals} but for the number of viable subhalos per isolated halo. The left-hand column use snapshots uniformly spaced in $a$, while the right-hand column has snapshots uniformly spaced in $\ln(a)$. Rows indicate results for $z=0$, 1 and 3 (from top to bottom).}
 \label{fig:convergenceNumberHiZ}
\end{figure*}

Figure~\ref{fig:convergenceNumberHiZ} shows convergence in the number of viable subhalos per isolated halo at different redshifts. An ``isolated halo'' is one which is not a substructure within a larger halo and so corresponds to the type of halo that might be found by a friends-of-friends algorithm in an N-body simulation. By a ``viable subhalo'' we mean any subhalo (including the main subhalo which hosts the central galaxy) which at some point in the merger tree was an isolated halo and so would have had the opportunity to potentially accrete gas from the intergalactic medium and form a galaxy. Whether or not such a subhalo actually would form a galaxy depends on the baryonic physics. Here we are simply assuming that any halo which was never isolated definitely would not form a galaxy\footnote{Non-viable substructures could of course be detected in N-body simulations, using an appropriate substructure finding algorithm such as {\sc Subfind} \protect\citep{springel_populating_2001}. Therefore, the results in Fig.~\ref{fig:convergenceNumberHiZ} are not directly relevant to HOD models or abundance matching models that are based on subhalo finding algorithms. However, within the context of current semi-analytic models, such subhalos could not form a galaxy, and so these results are relevant to HOD models that are based on fits to results from semi-analytic models that used N-body-derived merger trees.}. Convergence at low redshift occurs at the same rate when timesteps spaced uniformly in $a$ or $\ln(a)$ are used, while at high redshift timesteps spaced uniformly in $\ln(a)$ give faster convergence. This simply reflects the relative density of timesteps at low and high redshifts under these two choices for timestep distribution. We find once again that 128 steps are sufficient to achieve convergence in the number of viable subhalos to better than 5\% in all cases. Using 64 steps can result in errors of up to 10--15\% at higher redshifts.

\subsubsection{Mass Functions and Star Formation Rates}

\begin{figure*}
 \begin{center}
  \begin{tabular}{cc}
   \includegraphics[height=60mm]{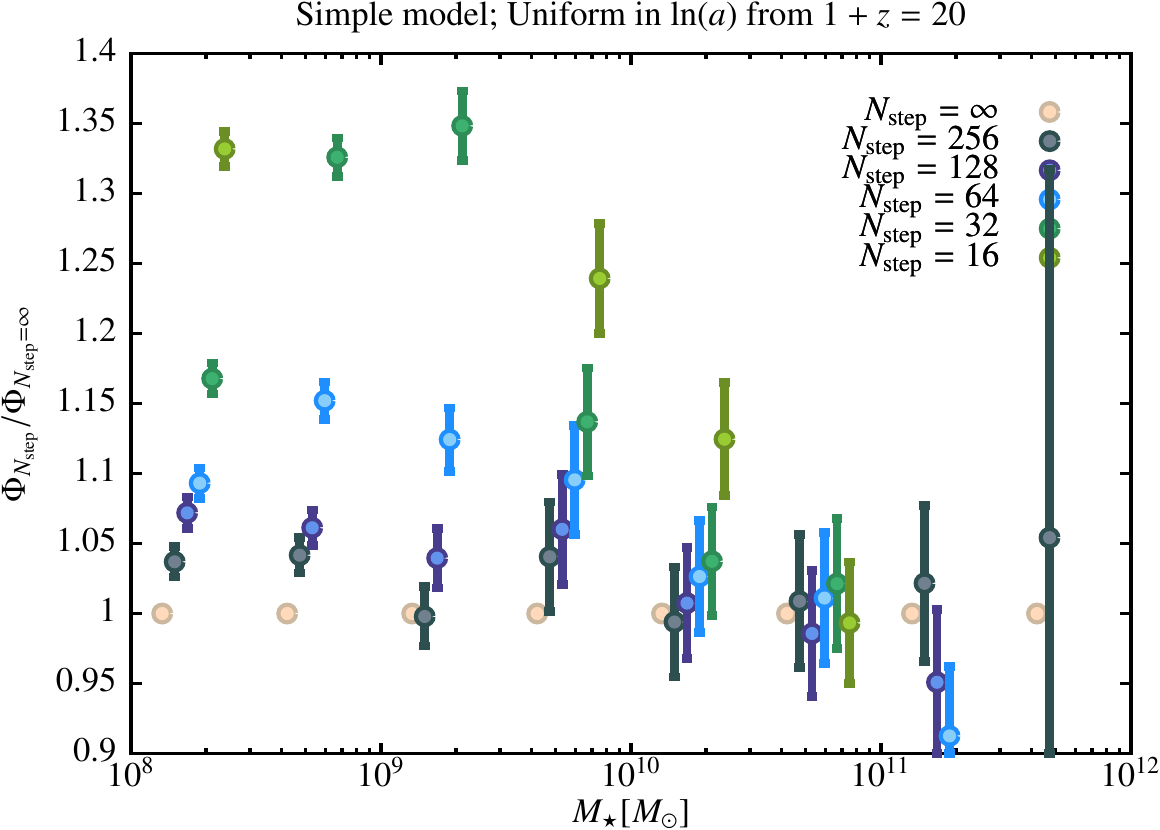} & \includegraphics[height=60mm]{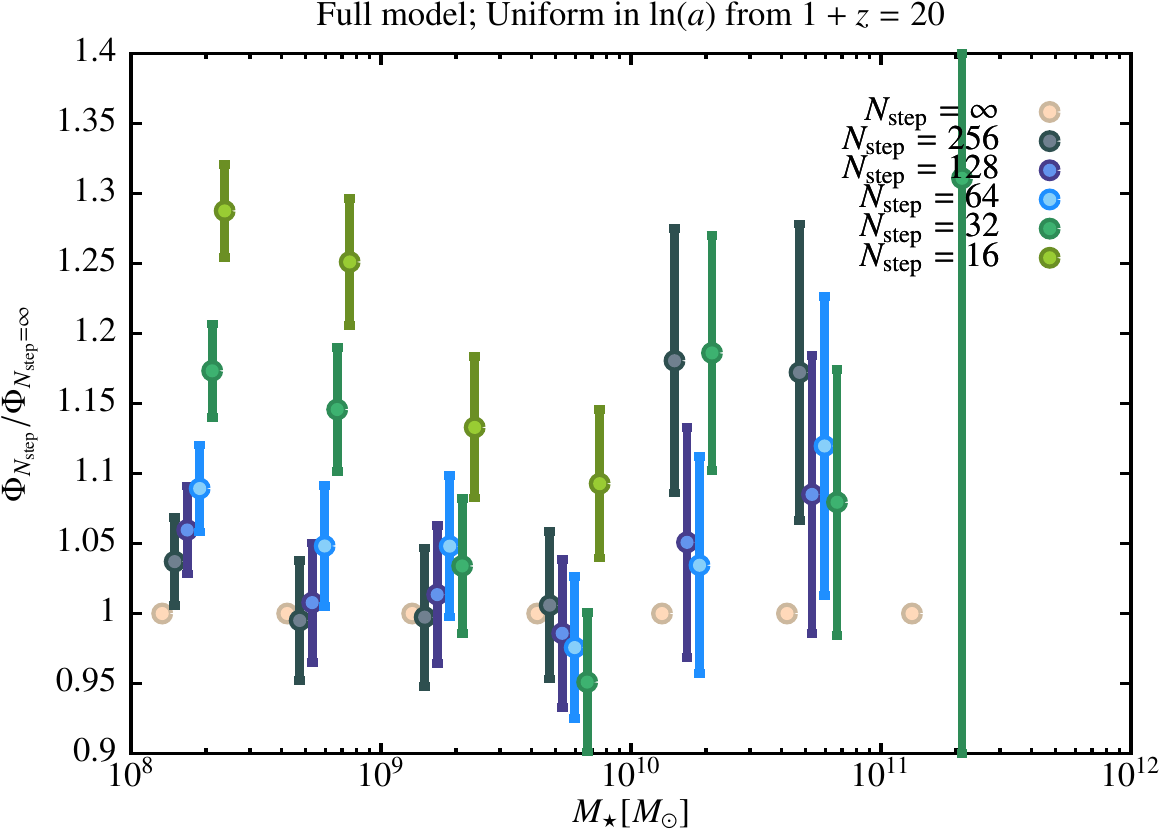} \\
   \includegraphics[height=60mm]{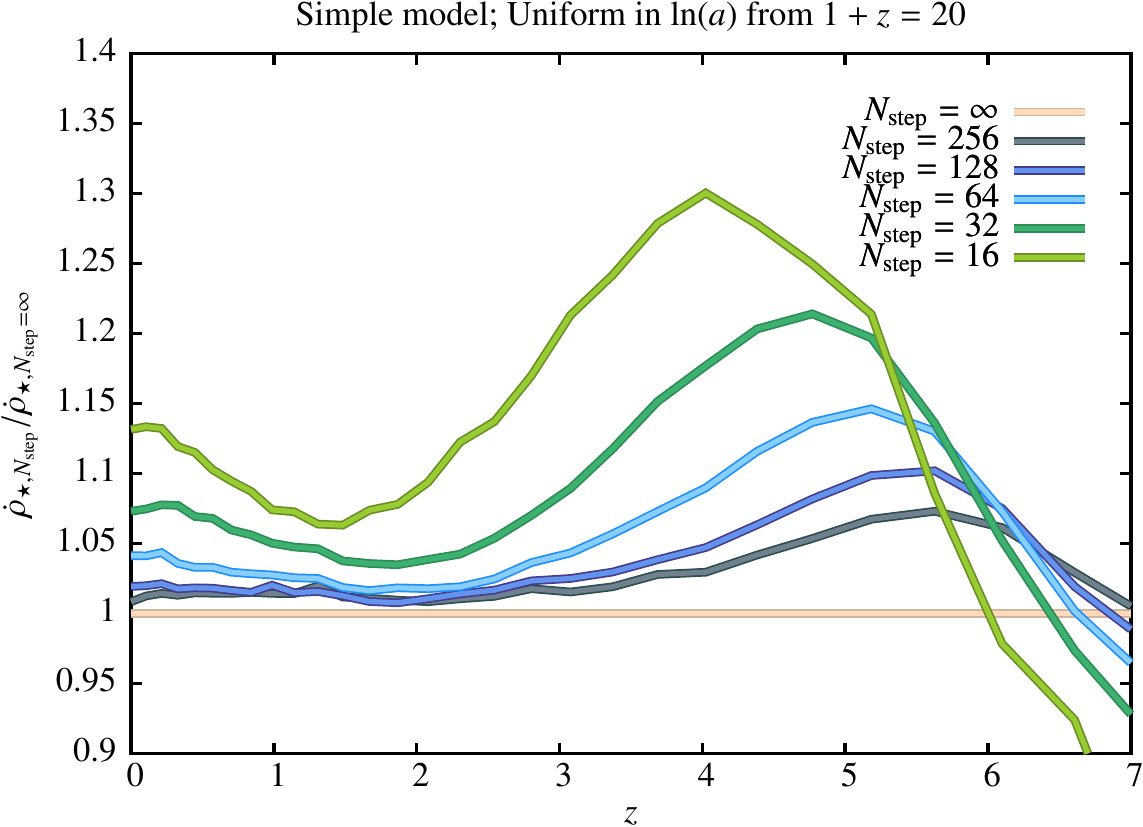} & \includegraphics[height=60mm]{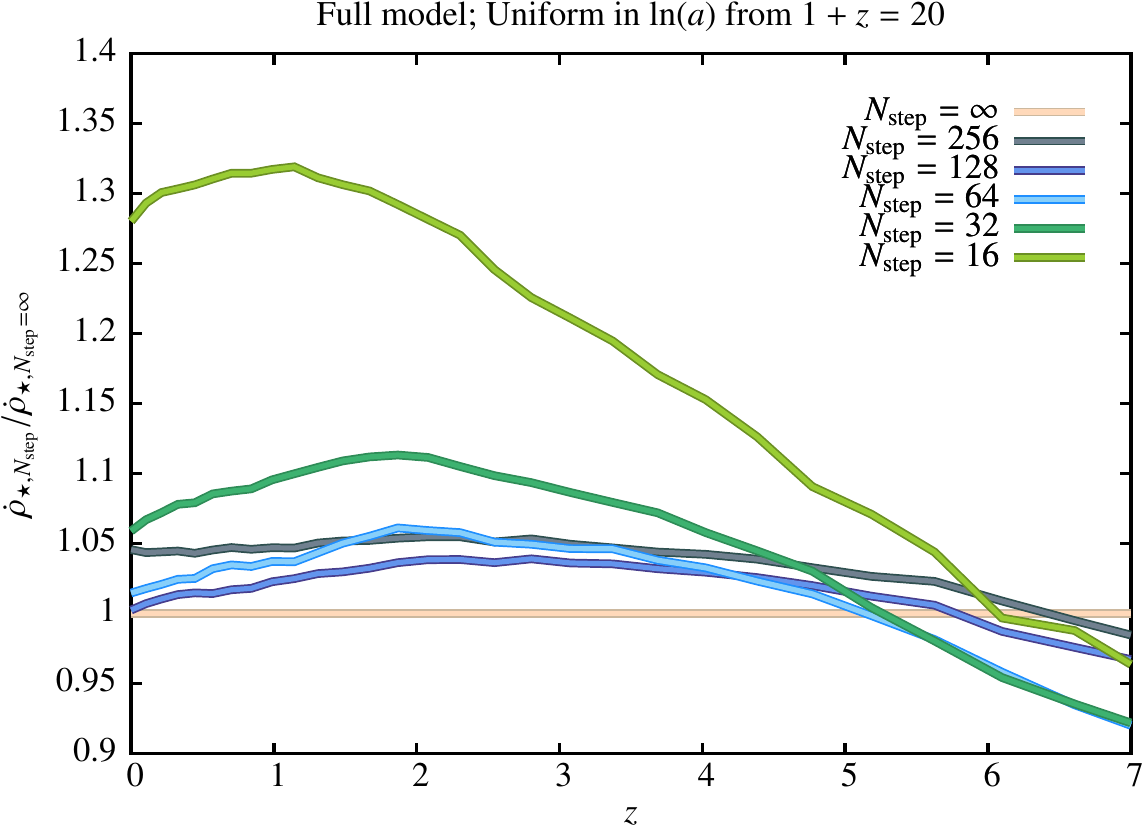} 
  \end{tabular}
 \end{center}
 \caption{Convergence with number of snapshots of the stellar mass function of galaxies at $z=0$ (upper panels), and the star formation history as a function of redshift (lower panels). Panels in the left column correspond to the simple model, while those in the right column correspond to the full model. In all cases snapshots are uniformly spaced in $\ln(a)$, and the earliest snapshot is at $1+z=20$ with the final snapshot at $1+z=1$. Symbol/line colour corresponds to the number of snapshots used. In the stellar mass function panels, error bars indicate the error on the mean mass due to the finite number of merger trees realized in each bin and points in each mass bin are given small horizontal offsets for clarity.}
 \label{fig:MFandSFR}
\end{figure*}

It is interesting to assess the convergence in statistics more closely related to observable quantities. In Fig.~\ref{fig:MFandSFR} we show convergence in the stellar mass function at $z=0$ (upper panels) and the volume density of star formation rate as a function of redshift (lower panels). 

The stellar mass functions show significant and systematic offsets as a function of $N_{\rm step}$. In both simple and full models convergence is worse for lower mass galaxies, with $N_{\rm step}=128$ in the simple model ($N_{\rm step}=64$ in the full model) being required to ensure better than 10\% convergence. 

For the star formation rate as a function of redshift we find significant and systematic offsets from the $N_{\rm step}=\infty$ case. In the simple model, using a finite number of timesteps results in an overestimate of the star formation rate at $z=0$ with a larger overestimate at $z=4$--$6$. Using $N_{\rm step}=128$ ensures convergence to better than 10\% across all redshifts. In the full model, convergence is more, rapid with even $N_{\rm step}=32$ getting close to 10\% or better convergence at all redshifts considered. 

\subsubsection{Major Merger Times}

\begin{figure*}
 \begin{center}
  \begin{tabular}{cc}
   \includegraphics[height=60mm]{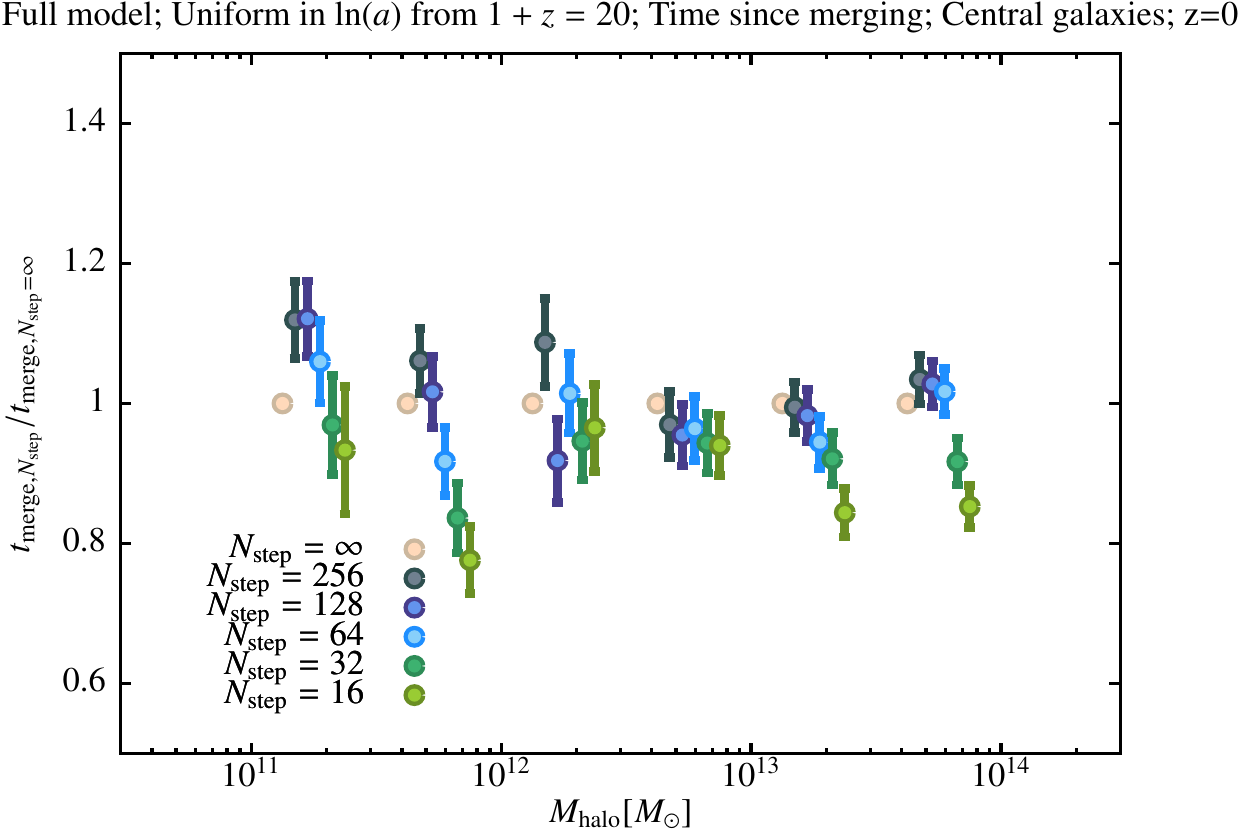} & \includegraphics[height=60mm]{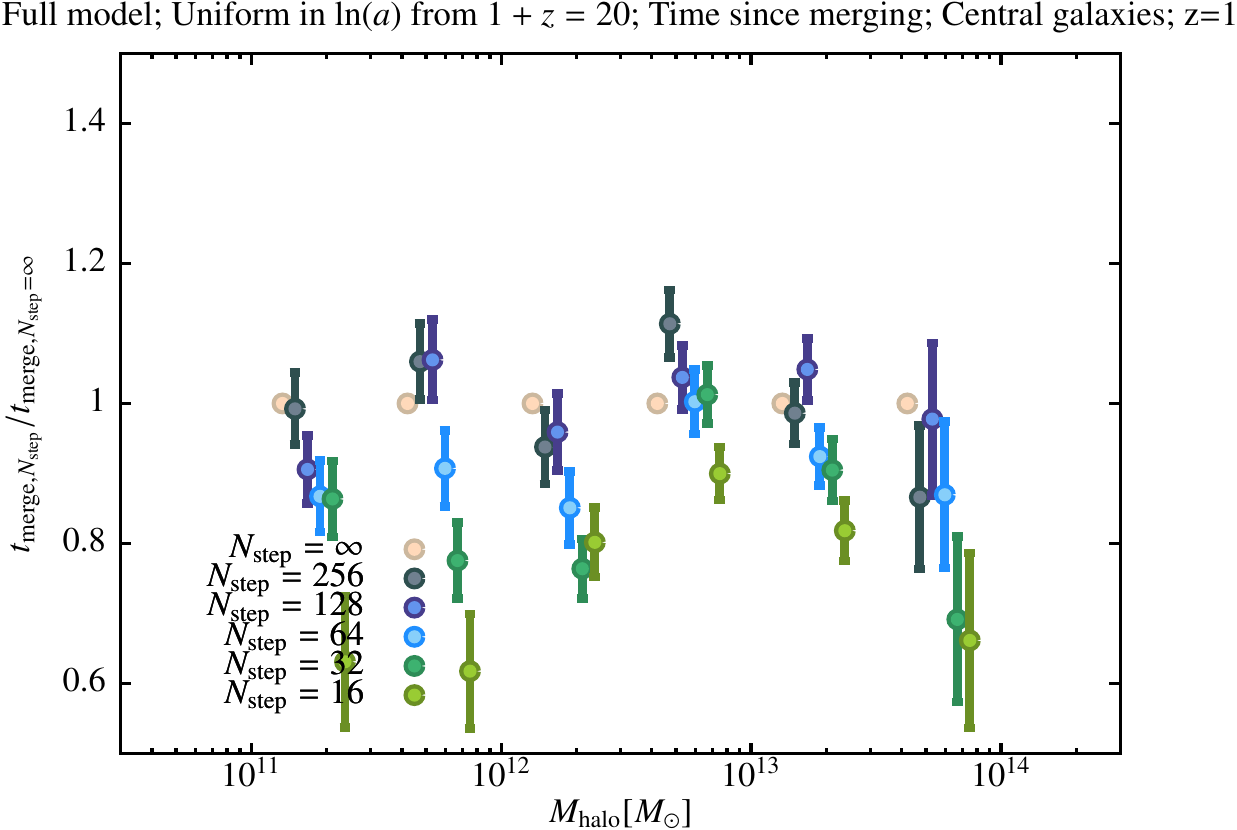}
  \end{tabular}
 \end{center}
 \caption{Convergence with number of snapshots of the times since the last major merger for central galaxies at $z=0$. The left panel corresponds to the simple model, while the right panel corresponds to the full model. Snapshots are uniformly spaced in $\ln(a)$, and the earliest snapshot is at $1+z=20$ with the final snapshot at $1+z=1$. Symbol colour corresponds to the number of snapshots used. Error bars indicate the error on the mean mass due to the finite number of merger trees realized in each bin. Points in each mass bin are given small horizontal offsets for clarity.}
 \label{fig:majorMergerTimes}
\end{figure*}

Finally, we examine the convergence in the distribution of merging times. Specifically, we consider the time since the last major merger experienced by central galaxies as a function of their halo mass. We compute the mean of this distribution, excluding any central galaxies which never experienced a major merger. Figure~\ref{fig:majorMergerTimes} shows the resulting convergence in this quantity as a function of $N_{\rm step}$ at $z=0$ (left panel) and $z=1$ (right panel) for the full model. For low values of $N_{\rm step}$ offsets from the true value are clearly seen, but in each case $N_{\rm step}=64$ is sufficient to achieve a converged answer.

\section{Conclusions}\label{sec:Conclusions}

It has become common practice to extract histories of the hierarchical merging process (``merger trees'') from cosmological N-body simulations and to use these as inputs to semi-analytic models of galaxy formation. Previously, there has been little consideration of how the temporal resolution of these trees affect the properties of the resulting galaxies. In this work we performed a convergence study using the \glc\ toolkit. 

We find that 128 snapshots spaced uniformly in the logarithm of expansion factor (or, almost equivalently, in the logarithm of the critical overdensity for collapse) provide good ($\sim 10\%$) convergence in galaxy stellar and total masses, the number of viable subhalos (i.e. those which have progenitors in the merger tree that are isolated, non-substructure halos), distributions of merger times, in stellar mass functions at $z=0$ and in the volume density of star formation rate as a function of redshift. Smaller numbers of snapshots lead to rapidly diverging results and should be avoided. We also considered snapshots spaced uniformly in expansion factor. We find that no substantial difference in the number of timesteps required to reach a given degree of convergence using this distribution of snapshots. This convergence is obtained for mean quantities averaged over large samples of galaxies---the full model in particular shows significant variance for individual galaxies even when using very large numbers of snapshots.

Our results should provide guidance as to how many snapshots should ideally be stored from future N-body simulations to ensure that the resulting temporally sparse merger trees do not overly limit the accuracy of subsequent galaxy formation calculations. Our results are for a specific set of cosmological parameters and tree mass resolution, in addition to being for a specific implementation of baryonic physics. The rate of convergence plausibly depends on all of these factors, and will likely differ for galaxy properties other than those considered here. Since \glc\ is freely available as an open source project\footnote{\protect\glc\ can be downloaded from \href{http://sites.google.com/site/galacticusmodel}{{\tt http://sites.google.com/site/galacticusmodel}}.}, it is relatively easy for anyone to repeat the analysis performed here for a specific set of simulation parameters and galaxy properties. Our results were obtained with v0.9.0.r491 of \glc\ and we have made the input parameters available online\footnote{The input parameter files and scripts to construct plots of the results can be downloaded from \href{http://www.ctcp.caltech.edu/galacticus/parameters/dmTreeConvergence.tar.bz2}{{\tt http://www.ctcp.caltech.edu/galacticus/parameters\\/dmTreeConvergence.tar.bz2}}.}.

Increasing mass resolution in simulations implies that merger trees contain more information. However, for this information to be folded into semi-analytic model predictions, trees must be built with finer time-stepping. Thus, increasing mass resolution would imply increasing both the size of each snapshot and the number of such snapshots, thereby causing a ``data tsunami''. It would then be recommendable for large high resolution simulations to be post-processed on-the-fly, writing at finely spaced times only (sub)halo catalogues instead of the entire snapshot.

Recent interest in exploring and constraining the parameter space of semi-analytic galaxy formation models \citep{henriques_monte_2009,bower_parameter_2010,lu_bayesian_2010} makes it crucial to understand and control numerical inaccuracies in such codes. Otherwise, quantitative constraints on model parameters will be subject to unknown systematic biases. While many uncertainties remain in our understanding of the physics of galaxy formation, it is important to ensure that numerical results are converged. Considerations such as those described here should become a standard part of any galaxy formation study.

\section*{Acknowledgments}

AJB acknowledges the support of the Gordon \& Betty Moore Foundation. SB has been partially supported by the PD51 INFN grant and by the PRIN-INAF grant ``Towards an Italian Network for Computational Cosmology''. GDL acknowledges financial support from the European Research Council under the European Community's Seventh Framework Programme (FP7/2007-2013)/ERC grant agreement n. 202781. MBK acknowledges support from the Southern California Center for Galaxy Evolution, a multi-campus research program funded by the University of California Office of Research.
We thank the KITP, Santa Barbara, where this work was begun, for their hospitality. This work made extensive use of Amazon's Elastic Compute Cloud through a generous grant from the Amazon in Education program.

\bibliographystyle{mn2e}
\bibliography{dmTreeConvergenceReduced}

\end{document}